\documentclass[3p]{elsarticle}
\usepackage{graphicx}
\usepackage{dcolumn}
\usepackage{bm}
\usepackage{lineno}
\usepackage{epsfig}
\usepackage{amsmath}
\usepackage{caption}
\usepackage{subfigure}

\newcommand{\no}[1]{}

\makeatletter
\renewcommand\p@subfigure{\thefigure}
\makeatother

\begin{document}

\begin{frontmatter}

\title{ Simulation of flux expulsion and associated dynamics in a two-dimensional magnetohydrodynamic channel flow }

\author{Vinodh Bandaru}
\author{Julian Pracht}
\author{Thomas Boeck}
\author{J\"{o}rg Schumacher}
\address{Institut f\"{u}r Thermo- und Fluiddynamik, Technische Universit\"{a}t Ilmenau, Postfach 100565, 98684 Ilmenau, Germany}

\begin{abstract}
We consider a plane channel flow of an electrically conducting fluid which is driven by a mean pressure gradient in the presence of an applied magnetic field that is streamwise periodic with zero mean. 
Magnetic flux expulsion and the associated bifurcation in such a configuration is explored using direct numerical simulations (DNS). The structure of the flow
and magnetic fields in the Hartmann regime (where the dominant balance is through Lorentz forces) and the Poiseuille regime (where viscous effects play a significant role) are studied and detailed comparisons to the existing one-dimensional model of Kamkar and Moffatt (J. Fluid. Mech.,
Vol.90, pp 107-122, 1982)
are drawn to evaluate the validity of the model. Comparisons show good agreement of the model with DNS in the Hartmann regime, but significant diferences arising in the Poiseuille regime
when non-linear effects become important. The effects of various parameters like the magnetic Reynolds number, imposed field wavenumber etc. on the bifurcation of the 
flow are studied. Magnetic field line reconnections occuring during the dynamic runaway reveal a specific two-step pattern that leads to the gradual expulsion of flux in the core 
region.
\end{abstract}


\end{frontmatter}

\section{Introduction}
\label{intro}
Interaction of electrically conducting flows with magnetic fields which forms the basis of the subject of magnetohydrodynamics (MHD), occur commonly in nature and are frequently utilized in 
industrial processes. Naturally they occur in astrophysical phenomena like the formation of stars and galaxies, phenomena in the Sun like sunspots and solar flares, and in the 
generation of terrestrial magnetic field in the core of the Earth by the dynamo action \cite{Moffat-book}. In industrial processes like continuous casting of steel and aluminium, material processing 
and plasma confinement in nuclear fusion applications, magnetic fields are applied advantageously for flow manipulation and control \cite{Cukierski:2008,Smolentsev:2008,Davidson:1999}. 
In all MHD flows, the magnetic field affects the flow through Lorentz forces, but the back reaction of the flow on the magnetic field (that leads to the bending of field lines) 
depends on a parameter called the magnetic Reynolds number defined as,
\begin{equation}
R_{m} = \frac{UL}{\lambda},
\end{equation}
where $U$ and $L$ are the charateristic velocity and length scales in the flow and $\lambda$ is the magnetic diffusivity of the fluid given by 
$\lambda = \left( \mu_{0}\sigma\right)^{-1}$, $\mu_{0}$ and $\sigma$ being the magnetic permeability of free space and the electrical conductivity of the fluid respectively. 
Significant bending of magnetic field lines occur only when $R_{m}\sim1$ and larger. Flows considered in this paper fall into this category.  

An interesting feature of MHD flows at high magnetic Reynolds numbers is the expulsion of magnetic flux that typically occurs under the imposition of an electrically conducting fluid flow with closed streamlines. This follows 
from an analogy of the well known Prandtl-Batchelor theorem \cite{Batchelor-book} in classical hydrodynamics. The kinematic problem of magnetic flux expulsion under rotation has been 
extensively studied during the sixties (see for e.g. \cite{Parker:1963,Parker:1966,Weiss:1966}) in the context of astrophysics. That flux expulsion also persists in the dynamic regime
was pointed out by Galloway et al. \cite{Galloway:1978} and was followed by further analysis of the dynamic effects of flux ropes in Rayleigh-B\'{e}nard magnetoconvection 
\cite{Proctor:1979}. 

An important aspect of the dynamic behavior associated with flux expulsion is the `runaway' effect, which can be explained as follows. When magnetic field lines start to get expelled
in a region of the flow, there is a decrease in the Lorentz forces that opposes the mean flow leading to an acceleration of the flow in that region. This in turn leads to further expulsion of 
magnetic flux and subsequently results 
in a cascading effect of flow acceleration and flux expulsion, wherein dissipative forces like viscosity ultimately balances the driving force, leading to a steady state. This effect can play a significant role in 
in the performance of electromagnetic pumps that are used to pump liquid metal. Early analytical studies of this phenomenon by Gimblett et al. \cite{Gimblett:1979} 
using rotating cylindrical and spherical solid bodies under an applied normal magnetic field showed an associated Thom cusp catastrophe and hysteresis effect. Such behaviour was seen 
further in the fluid context by Moffatt \cite{Moffat:1980}. 

However, flux expulsion can also happen in flow configurations without closed streamlines, if the imposed magnetic field is non-uniform and periodic in the mean flow direction. A
particularly interesting configuration is that of a plane channel flow driven by a mean pressure gradient with an imposed sinusoidal magnetic field that was analysed by Kamkar and
Moffatt \cite{Kamkar:1982} which will be denoted as KM82 hereafter. In their study, the interaction of the flow and magnetic fields was described by simplified one-dimensional 
model equations. Steady state solutions were obtained 
from which two different flow regimes were 
identified, namely the Hartmann and Poiseuille regimes and the location of the bifurcation leading to the transition between these two regimes was computed. However, various
simplifications were assumed in that study. For example, the non-linear terms (and hence the Reynolds stress terms) in the Navier-Stokes equation and the variations along the 
streamwise direction were neglected. Although it enables one to obtain quick solutions, the approximate model can lead to significant loss of accuracy and underprediction/overprediction
of the jump that occurs during the bifurcation. The focus of our work is to perform 2D direct numerical simulations of the problem similar to KM82 with a 
twofold purpose. On one hand, it helps validate the 1D model
predictions at the steady state and quantify the differences arising out of the simplifications of the model. On the other hand, the presence of non-linearities can result in 
time-dependent solutions for the flow and magnetic fields in both the regimes in the final state. 

The paper is organized as follows. Section~\ref{sec:psetup} describes the problem setup and the full governing equations along with a brief overview of the KM82 model. This is
followed by the details of the numerical procedure in 
section~\ref{sec:numerics}. In section~\ref{sec:results}, 
numerical results of the DNS are presented and compared to the predictions of the model, including the effect of various parameters on the charateristics of the bifurcation,
followed by conclusions in section~\ref{sec:conclusions}.

\section{Problem setup and governing equations}
\label{sec:psetup}
\subsection{Problem formulation and full governing equations}
\begin{figure}[!h]
  \includegraphics[width=1.0\textwidth,trim=10mm 10mm 15mm 120mm,clip]{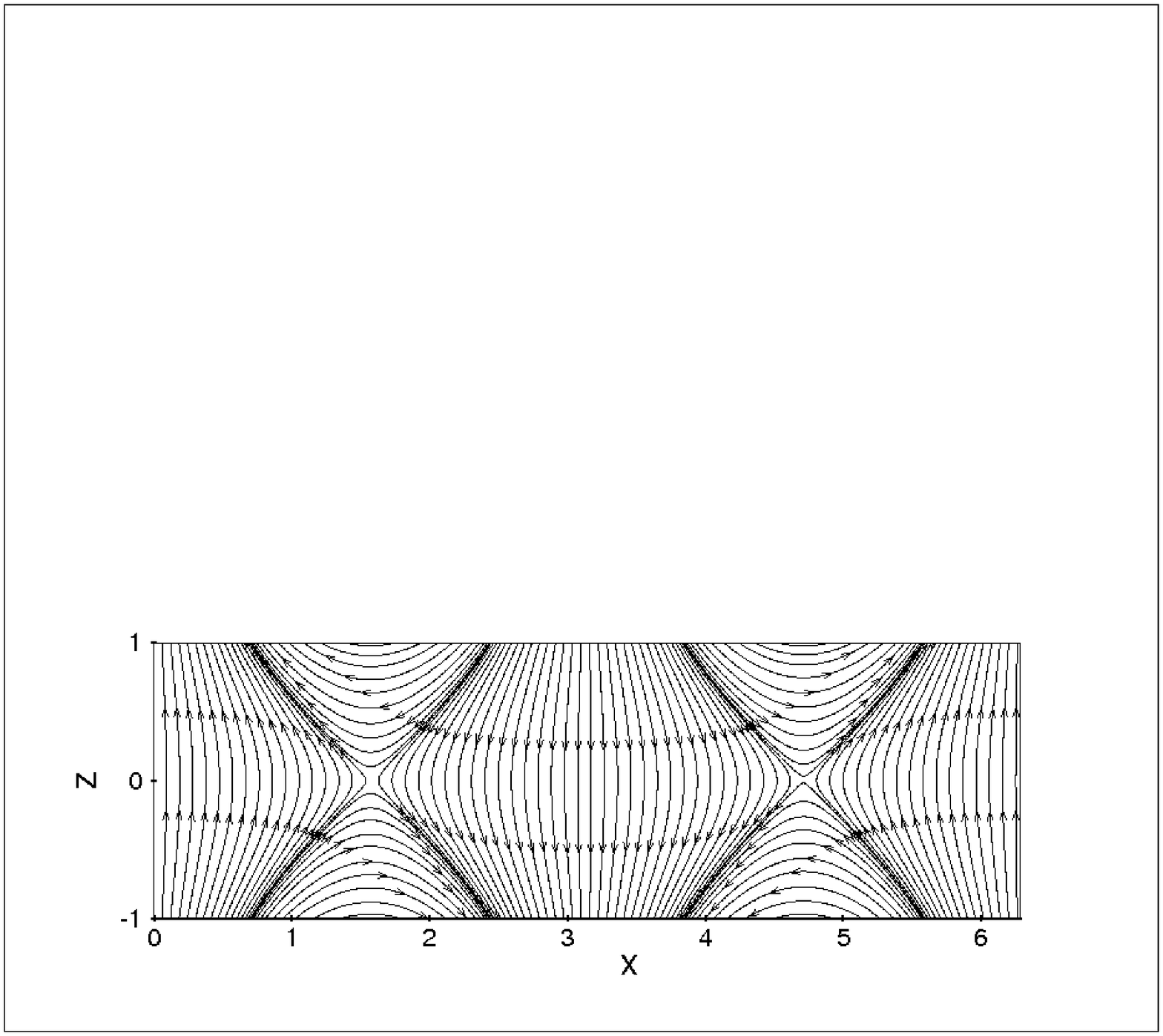}
\caption{Field lines of the imposed magnetic field $\bm{b}_{0}$ as given by equation \eqref{b0}. Two magnetic X-points at ($x=\pi/2, z=0$) and ($x=3\pi/2, z=0$) can be observed
 at the centerline.}
\label{fig:b0}       
\end{figure}

We consider the two-dimensional incompressible flow of an electrically conducting fluid (e.g. a liquid metal) driven by a mean pressure gradient in a straight rectangular channel. Periodicity is assumed along the 
streamwise direction $x$ and the wall normal direction is denoted by $z$. A magnetic field $\bm{B}_{0}(x,z)=\bm{B}(x,z,t=0)$ (generated by electric current or magnet sources outside the
channel) with a prescribed wall normal component $B_{0z}=cos(kx)$ with 
a wavenumber $k$, is imposed on the flow. Such a periodic magnetic field can be ideally produced for e.g. by magnets distributed on the channel walls with alternating north and south
poles in the streamwise direction (see \cite{Kamkar:1982}).
The action of the flow on the magnetic field
generates plane-normal electric current densities $\bm{J}=(0,J(x,z),0)$ in the flow which leads to the generation of a secondary magnetic field and Lorentz forces that 
affect the flow. The imposed magnetic field which is divergence-free can be expressed as $\bm {B}_{0}=\nabla \times \bm{A}_{0}$, where $\bm{A}_{0} = \left(0,A_{0}(x,z),0\right) $ is the magnetic vector 
potential. Choosing the scales of half channel height $L$ for the length, the maximum value of the imposed magnetic field $B_{0}$ for the magnetic field, $B_{0}L$ for 
the vector potential and applying the curl-free condition on $\bm {B_{0}}$, we get in the non-dimensional form,
\begin{equation}
\label{laplacea} \frac{\partial^2 a_{0}}{\partial x^2} + \frac{\partial^2 a_{0}}{\partial z^2} = 0 \textrm{,}  
\end{equation}
with the boundary conditions 
\begin{equation}
\label{bca} \frac{\partial a_{0}}{\partial x} = cos(\kappa x) \hskip2mm \textrm{on} \hskip2mm z=\pm1 \hskip1mm \textrm{;} \hskip2mm a_{0}\left(0,z\right) = a_{0}\left(l_{x},z\right)
\end{equation}
where the symbol $\kappa=kL$ represents the normalized wavenumber, $l_{x}$ is the non-dimensional streamwise length of the channel and $a_{0}$ is the normalized vector potential.
Solution of equation \eqref{laplacea} using seperation of variables yields
\begin{equation}
\label{Aeqn} a_{0} = \frac{1}{\kappa} \frac {sin(\kappa x)cosh(\kappa z)}{cosh(\kappa)} \textrm{.}  
\end{equation}
This leads to the form of the non-dimensional initial (or imposed) magnetic field $\bm{b}_{0}$ as 
\begin{equation}
\label{b0} \bm{b}_{0} =  -\frac {sin(\kappa x)sinh(\kappa z)}{cosh(\kappa)} \bm{i} + \frac {cos(\kappa x)cosh(\kappa z)}{cosh(\kappa)} \bm{k} \textrm{,}
\end{equation}
the field lines of which are shown in Fig.~\ref{fig:b0}. Here $\bm{i}$ and $\bm{k}$ refers to the unit vectors in the streamwise ($x$) and wall-normal ($z$) directions respectively.

The physics of the problem is governed by the Navier-Stokes equation for the momentum balance including the additional source term representing the Lorent force (body force) produced
by the induced electric currents and the induction equation for magnetic field transport along with the constraints of 
mass conservation (continuity) and solenoidality of the magnetic field. The pressure gradient is decomposed as $\nabla P_{T} =-\rho G \bm{i} + \nabla P$, where $\nabla P_{T}$
represents the cumulative pressure gradient and $-\rho G$ the
constant mean pressure gradient that is applied in the streamwise direction. Non-dimensionalizing using the scales
 $\lambda/L^2k$, $\lambda/L^2kG$, $\rho GL$ and $\sigma \lambda B_{0}/L^2 k$ for velocity,
time, pressure and the current densities respectively, and denoting all non-dimensional variables by small letters, the system of governing equations can be written as

\begin{eqnarray}
\label{navierstokes} \hskip-15mm& & \frac{\partial \bm{v}}{\partial t} + \frac{1}{\beta\kappa}
(\bm{v}\cdot \nabla)\bm{v} = 1 - \nabla p +  \epsilon \nabla^2 \bm{v} + \frac{1}{Q}\left(\bm{j} \times \bm{b}\right),\no{MHDeqn}\\
\label{btransport} \hskip-15mm& & \frac{\partial \bm{b}}{\partial t} +
\frac{1}{\beta\kappa}(\bm{v}\cdot \nabla)\bm{b} = \frac{1}{\beta\kappa}(\bm{b}\cdot \nabla)\bm{v} + \frac{1}{\beta}\nabla^2
\bm{b},\no{MHDeqn}\\
\label{continuity} \hskip-15mm& & \nabla \cdot\bm{v}  =  0,\\
\label{divb} \hskip-15mm& & \nabla \cdot\bm{b}  =  0,\\
\label{amperelaw} \hskip-15mm& & \bm{j} = \kappa\left( \nabla \times\bm{b}\right) ,\\
\label{MHDeqn_bc} \hskip-15mm& & u = w = 0, \hskip1mm b_{z} = cos(\kappa x) \hskip1mm\mbox{on } z=\pm1 \hskip1mm; \quad
\textrm{$\bm{v}$, $\bm{b}$ periodic in $x$-direction} \no{MHDeqn_bc}
\end{eqnarray}
 with the no-slip and no penetration boundary conditions for the fluid velocity on the walls and periodicity assumed in the streamwise direction. Furthermore, the wall normal 
component of the magnetic field $b_{z}$ on the walls remain unchanged (equal to the imposed magnetic field, $b_{0z}$) and the streamwise 
component follows from the divergence-free condition \eqref{divb}. The parameters involved in the problem are 
\begin{equation}
\label{non-dim} \epsilon= \frac{\nu\lambda}{L^{4}kG} \textrm{,}\hskip7mm \beta= \frac{L^{4}kG}{\lambda^{2}} \textrm{,}\hskip7mm Q=\frac{\rho L^{2}kG}{\sigma \lambda B_{0}^2}\textrm{,}\hskip7mm \kappa=kL
\end{equation}
where $\nu$ and $\rho$ represent the kinematic viscosity and the mass density of the fluid respectively. The parameter $\beta$ represents the magnetic Reynolds number and the parameters
$\epsilon$, $Q$ can be regarded as the inverse of the hydrodynamic Reynolds number and the Stuart numbers respectively. All the fluid properties are assumed to be constant. The coupled
evolution of the velocity and magnetic fields is computed by solving the governing equations numerically, a brief summary of which is presented in the next section.

\subsection{One-dimensional approximate model of Kamkar and Moffatt (KM82)}
In view of the fact that the results of our DNS are compared with the 1D approximate model of KM82, we present a brief overview of their model here. The following assumptions 
have been made in the model
\begin{itemize}
\item The secondary magnetic field consists of only a single mode (wavenumber) along the streamwise direction, which is taken to be the wavenumber ($k$) of the applied 
magnetic field. To this effect, the vector potential $A$ of the magetic field $\bm{b}$ is expanded as 
\begin{equation}
\label{Aexpansion} A(x,z,t)= B_{0}k^{-1} \Re\left[ i f(z,t) e^{ikx}\right]
\end{equation}
where $\Re$ represents the real part and $f(z,t)$ is the dimensionless profile function.
\item The variation of dynamics in the streamwise direction is neglected and hence the mean ($x$-averaged) governing equations are considered.
\item The velocity fluctuations $\bm{v}'$ are small compared to the mean ($x$-averaged) streamwise velocity $U$, $|\bm{v}'|\ll U$ and hence the Reynolds stress terms in the momentum equation and the 
fluctuating parts of advection terms in the $A$-transport are neglected. This assumption is supposed to be valid when $\beta \ll Q^{2}$.
\end{itemize}
The assumptions stated above lead to the following mean governing equations for $U(z,t)$ and $f(z,t)$ in the non-dimensional form,

\begin{eqnarray}
\label{navierstokesk82} \hskip-15mm& & \frac{\partial U}{\partial t} = 1 - \frac{1}{2Q}\Re \left[ i f\left(\frac{\partial^2}{\partial z^2}-\kappa^2\right)f^{*}\right]   + \epsilon \frac{\partial^2U}{\partial z^2},\\
\label{btransportk82} \hskip-15mm& & \beta\frac{\partial {f}}{\partial t} + iUf = \left(\frac{\partial^2}{\partial z^2}-\kappa^2\right)f \\
\label{MHDeqn_bck83} \hskip-15mm& & U = 0,\hskip1mm f = 1 \hskip2mm\mbox{on } z=\pm1 \hskip1mm; 
\end{eqnarray}
which correspond to equations (2.33) and (2.38) in KM82, where $f^{*}$ is the complex conjugate of $f$. For later comparisons with the DNS results, we solve the above model equations
\eqref{navierstokesk82} to \eqref{MHDeqn_bck83} by a finite-difference method on a uniform grid.

\section{Numerical Procedure}
\label{sec:numerics}
\begin{figure}[!h]
\includegraphics[width=1.0\textwidth,trim=10mm 10mm 15mm 120mm,clip]{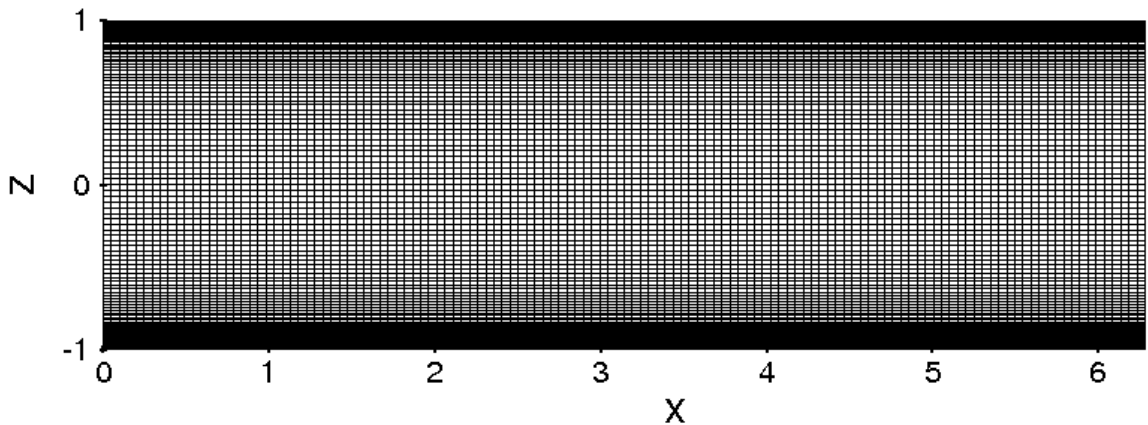}
\caption{Non-uniform structured grid in the $xz$-plane with $129\times129$ grid points and $S=2.2$.}
\label{fig:grid}     
\end{figure}

The numerical solution of the system \eqref{navierstokes} to \eqref{MHDeqn_bc} is obtained using a second order finite difference method. The domain is discretized into a rectangular Cartesian grid with
uniform grid spacing along the streamwise direction and a non-uniform stretched grid in the wall normal direction in order to resolve the thin Hartmann boundary layers near the walls,
that are typical of wall bounded MHD flows.
The stretched grid is obtained by a coordinate transformation from a uniform grid coordinate $\eta$ to the non-uniform grid coordinate $z$ through
\begin{equation}
\label{eq:gridstretch} z = L\frac{\tanh(S\eta)}{\tanh(S)} \hskip2mm \textrm{,} -1\leq \eta \leq 1 \hskip2mm \textrm{,}
\end{equation}
where $S$ represents the stretching factor. A typical grid used in our studies is shown in Fig.~\ref{fig:grid}.

The momentum equation \eqref{navierstokes} is integrated by the standard projection scheme where the non-linear, viscous and Lorentz force terms are treated explicitly by the 
Adams-Bashforth method to first compute an intermediate velocity field $\bm{v}^*$ which is then projected onto a solenoidal velocity field $\bm{v}^{n+1}$ at the new time level by 
a correction obtained from the solution of a Poisson equation for the pressure. A second order backward finite difference scheme is used for time discretization using the levels $n-1$, $n$,
$n+1$ where $n$ is the current time level. Details of the procedure to compute the velocity field can be found in Krasnov et al. \cite{Krasnov:2011}. For the magnetic field, the normal
component of the magnetic field $b_{z}$ is computed by using a semi-implicit procedure, wherein only the diffusive term is treated implicitly and the advective and field stretching 
terms are treated explicitly. This requires again the solution of a Poisson equation for $b_{z}$. The FORTRAN software package FISHPACK \cite{FISHPACK} has been used to solve the 
Poisson equations. Subsequently, the streamwise component $b_{x}$ is reconstructed from $b_{z}$ using equation \eqref{divb}, in order to satisfy the solenoidality of the magnetic field. 
Alternatively, it is possible to solve for the vector potential $A$ (since the problem is 2D, $A$ has only one component) and recover the magnetic field components $b_{x}$ and $b_{z}$
from $A$. The magnetic field so obtained is used to compute the current density $j$ which is required to evaluate the Lorentz force term in the Navier-Stokes equation. 
OpenMP parallelization has been used in performing the computations for the results presented in this paper. 

\section{Results and comparison}
\label{sec:results}
Starting with either an initial laminar velocity profile (with no streamwise variation) or fluid at rest ($\bm{v}=0$) and the imposed magnetic field $\bm{b_{0}}$, the governing equations are numerically integrated in time 
to obtain the final equilibrium states. All the computations have been performed for a streamwise domain length of one period, $l_x= 2\pi/\kappa $ on a $129\times129$ grid. A grid 
sensitivity study was performed which indicated that further increase in grid resolution does not improve the solution accuracy significantly, within the parameter space studied in
this paper. Depending on the velocity profiles of the final states, two regimes of flows are defined (as in KM82), namely the Hartmann regime and the Poiseuille 
regime.   

\begin{figure}[!h]
        \subfigure[]{
                \centering
                \includegraphics[width=0.5\textwidth]{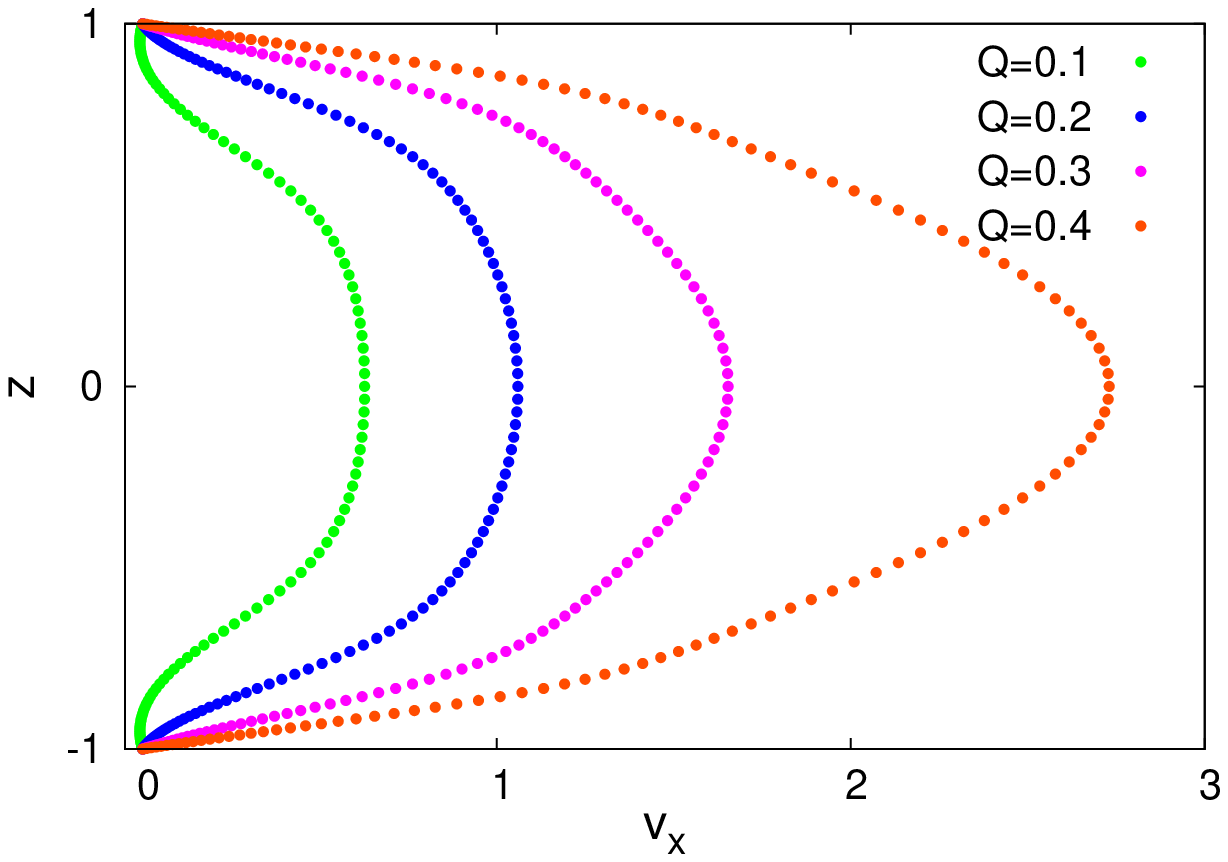}
                \label{fig:vxpiby2}
        }
        \subfigure[]{
                \centering
                \includegraphics[width=0.5\textwidth]{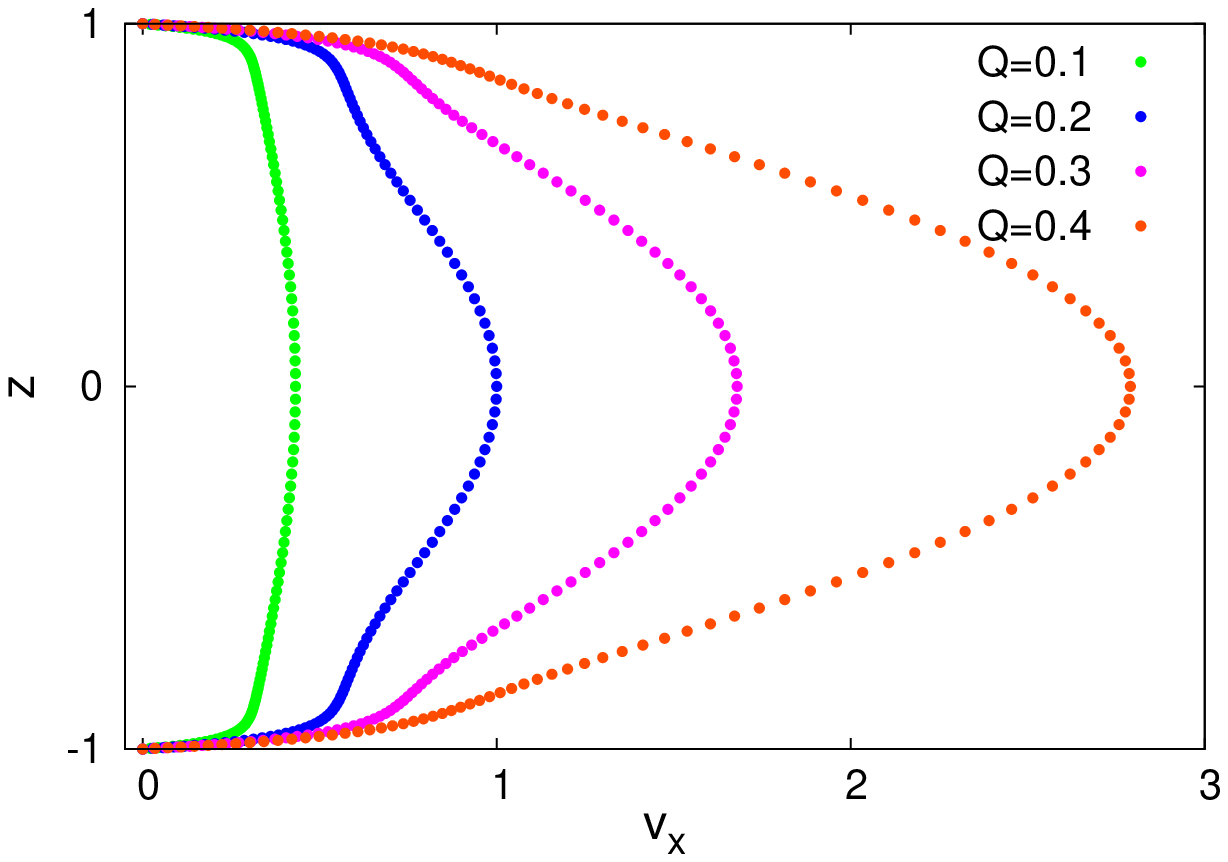}
                \label{fig:vxpi}
        }
\caption{ Steady state streamwise velocity profiles in the Hartmann regime at (a) $x=\pi/2$ and (b) $x=\pi$. Parameters are $\beta=1$, $\kappa=1$ and $\epsilon=5\times10^{-3}$.}
\label{fig:vx_profiles_hartmann}
\end{figure}
\begin{figure}[]
        \subfigure[]{
                \centering
                \includegraphics[width=1.0\textwidth,trim=5mm 60mm 10mm 60mm,clip]{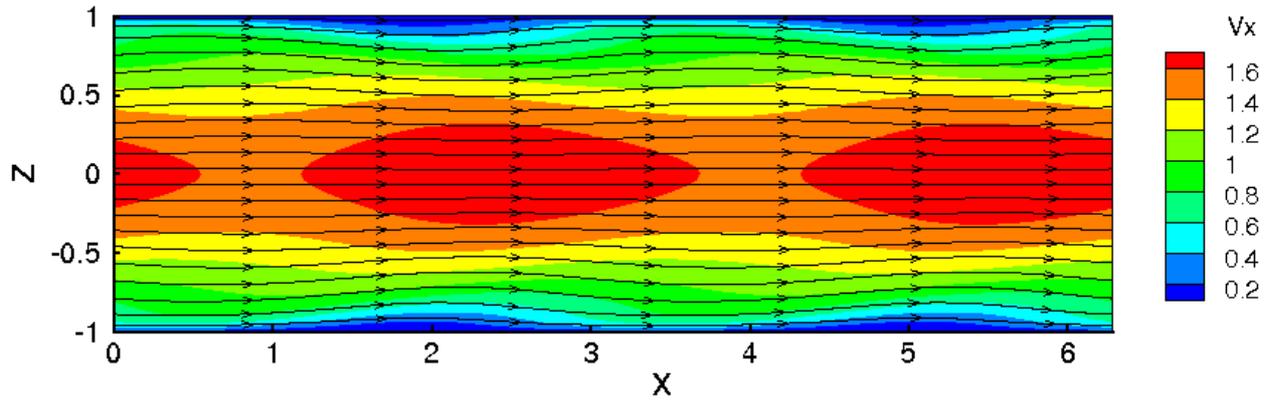}
                \label{fig:vxcontour_hartmann}
        }
        \subfigure[]{
                \centering
                \includegraphics[width=0.9\textwidth,trim=7mm 60mm 30mm 60mm,clip]{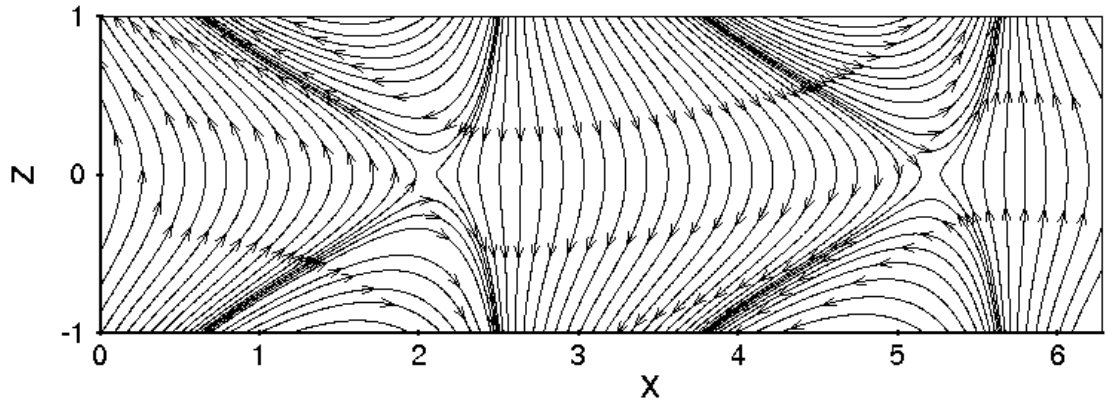}
                \label{fig:flines_hartmann}
        }
\caption{ Typical streamwise velocity and magnetic field configuration in the Hartmann regime. (a) Contours of streamwise velocity and (b) magnetic field lines in the steady state at $Q=0.3$, for $\beta=1$, $\kappa=1$ and $\epsilon=5\times10^{-3}$. We observe how
the magnetic X-points are slightly shifted by the flow.}
\label{fig:vx_contour_flines}
\end{figure}
The Hartmann regime is characterized by very steep velocity gradients in the boundary layers as compared to the core region. Pressure gradient in the core
is dominantly balanced by Lorentz forces whereas in the boundary layers it is a combination of viscous and Lorentz forces that balances the pressure gradient. Flows at relatively 
small $Q$ or high interaction parameter belong to this regime. In contrast, the Poiseuille regime typically demonstrates `Poiseuille-like' (parabolic) axial velocity profiles and
 is dominated by viscosity in the core region. Flows at relatively 
higher $Q$ belong to the Poiseuille regime. The final equilibrium state of the Hartmann
regime is observed to be steady in time unlike the Poiseuille regime where significant velocity fluctuations persist in final state. Transition between the two regimes occurs over a 
narrow band of $Q$ through a bifurcation that is a manifestation of the runaway effect. We now provide a very brief account of the nature of the solutions in these two regimes.

\subsection{Hartmann regime}
Typical axial velocity profiles in the Hartmann regime are shown in Fig.~\ref{fig:vx_profiles_hartmann} at two different axial locations of the channel, $x=\frac{\pi}{2}$ and $x=\pi$ and
at various values of the parameter $Q$. 
These two locations correspond to the streamwise extreme values of the imposed magnetic field $\bm{b_{0}}$. It can be observed from Fig.~\ref{fig:vx_profiles_hartmann} that higher axial
velocities (or flow acceleration from the initial state) are observed with increase in $Q$ and the profiles in the boundary layers at $x=\pi$ look more `Hartmann-like' than at 
$x=\frac{\pi}{2}$ due to the pronounced effect of the wall normal component $b_{0z}$. The distribution of axial velocity $v_{x}$ in the domain as shown in Fig.~\ref{fig:vxcontour_hartmann}
(along with the velocity streamlines) clearly indicates the laminar nature of the flow in the Hartmann regime. Advection of the magnetic field can be observed from the corresponding 
field lines shown in Fig.~\ref{fig:flines_hartmann} that indicate only a slight bending of the field lines. As is clear
from the velocity field, no significant events of 
\begin{figure}[!h]
        \subfigure[]{
                \includegraphics[width=0.48\textwidth,trim=2mm 50mm 2mm 40mm,clip]{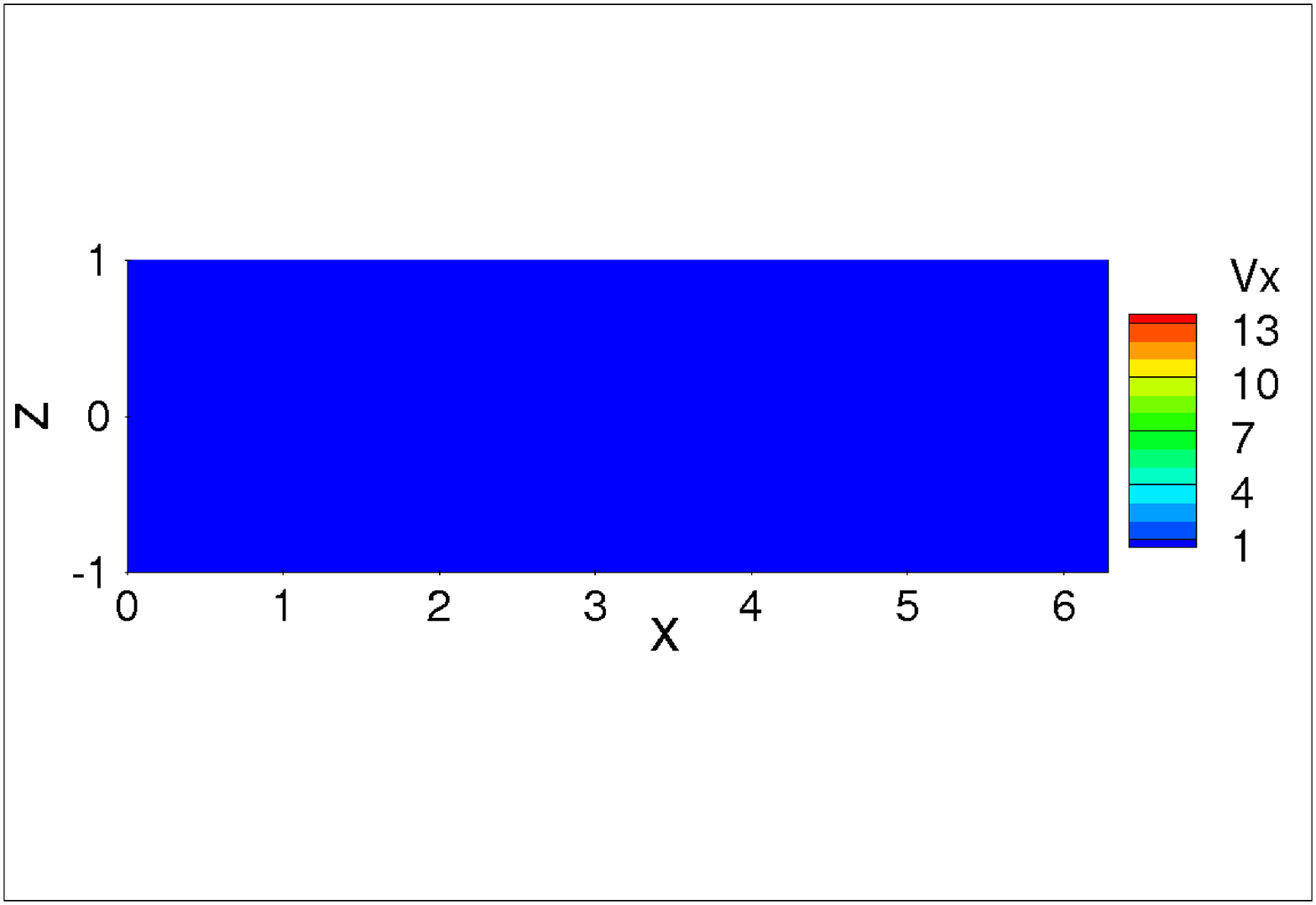}
                \includegraphics[width=0.48\textwidth,trim=2mm 50mm 2mm 40mm,clip]{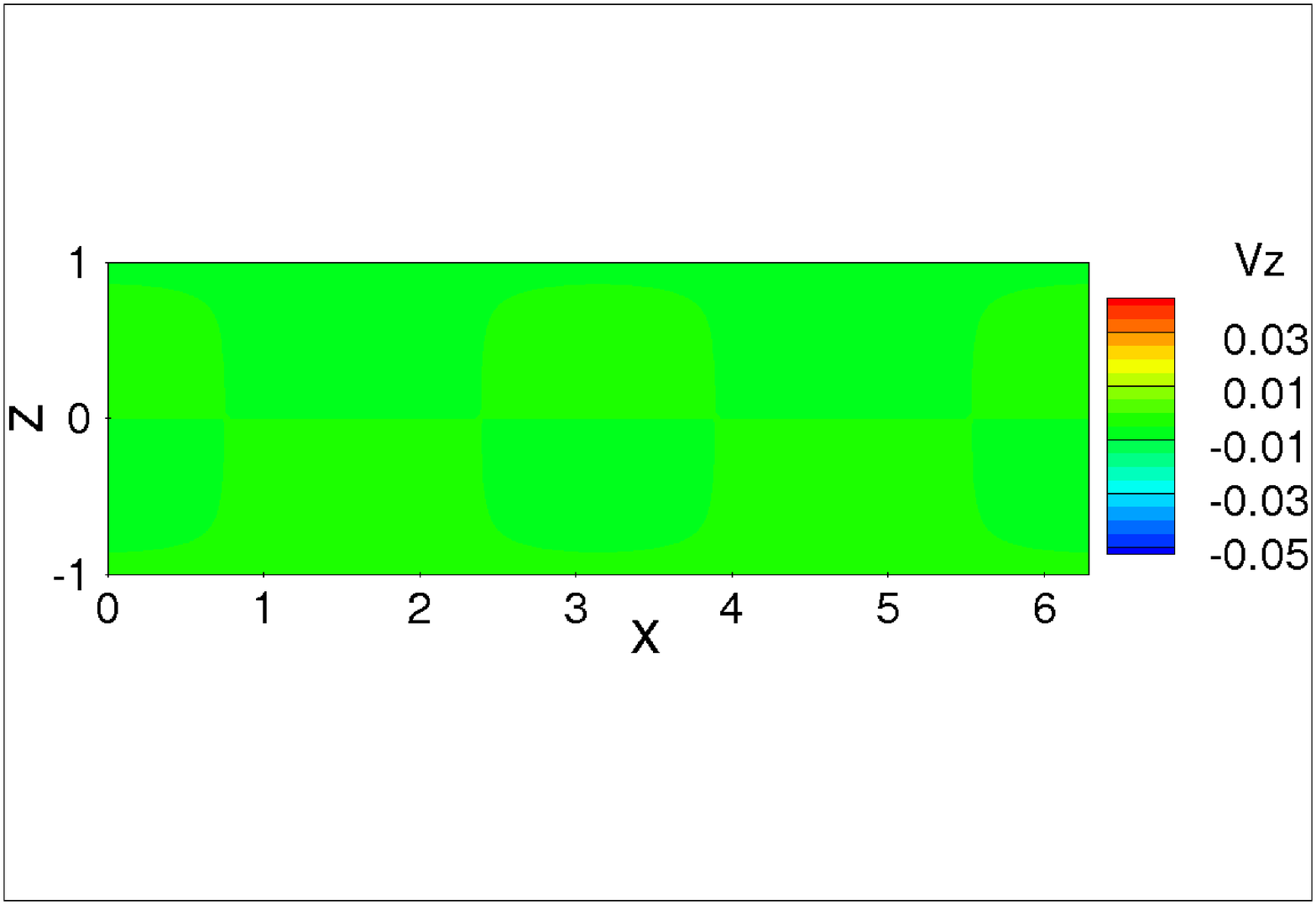}
                \label{fig:t0vxy}
        }
        \subfigure[]{
                \includegraphics[width=0.48\textwidth,trim=2mm 50mm 2mm 40mm,clip]{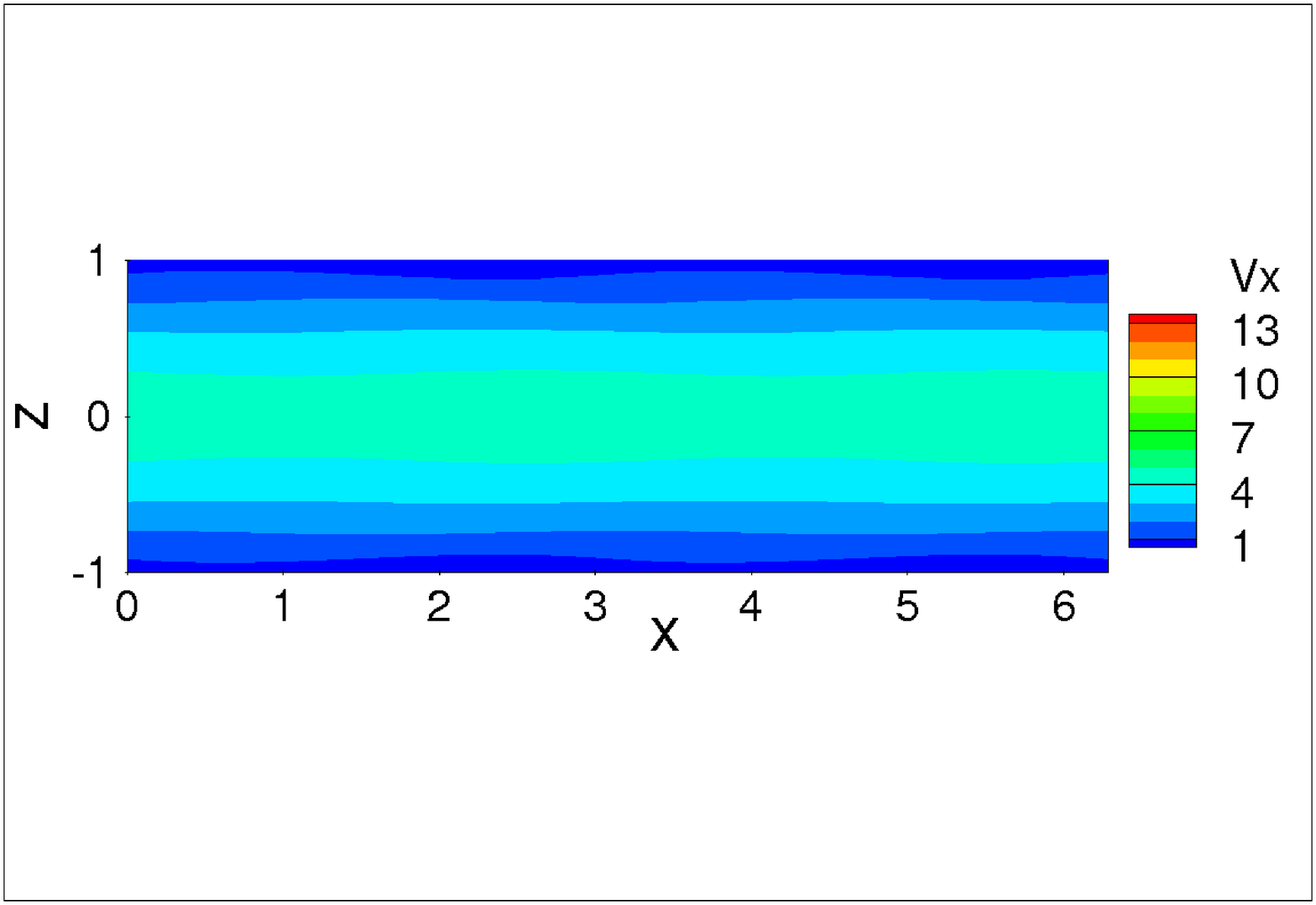}
                \includegraphics[width=0.48\textwidth,trim=2mm 50mm 2mm 40mm,clip]{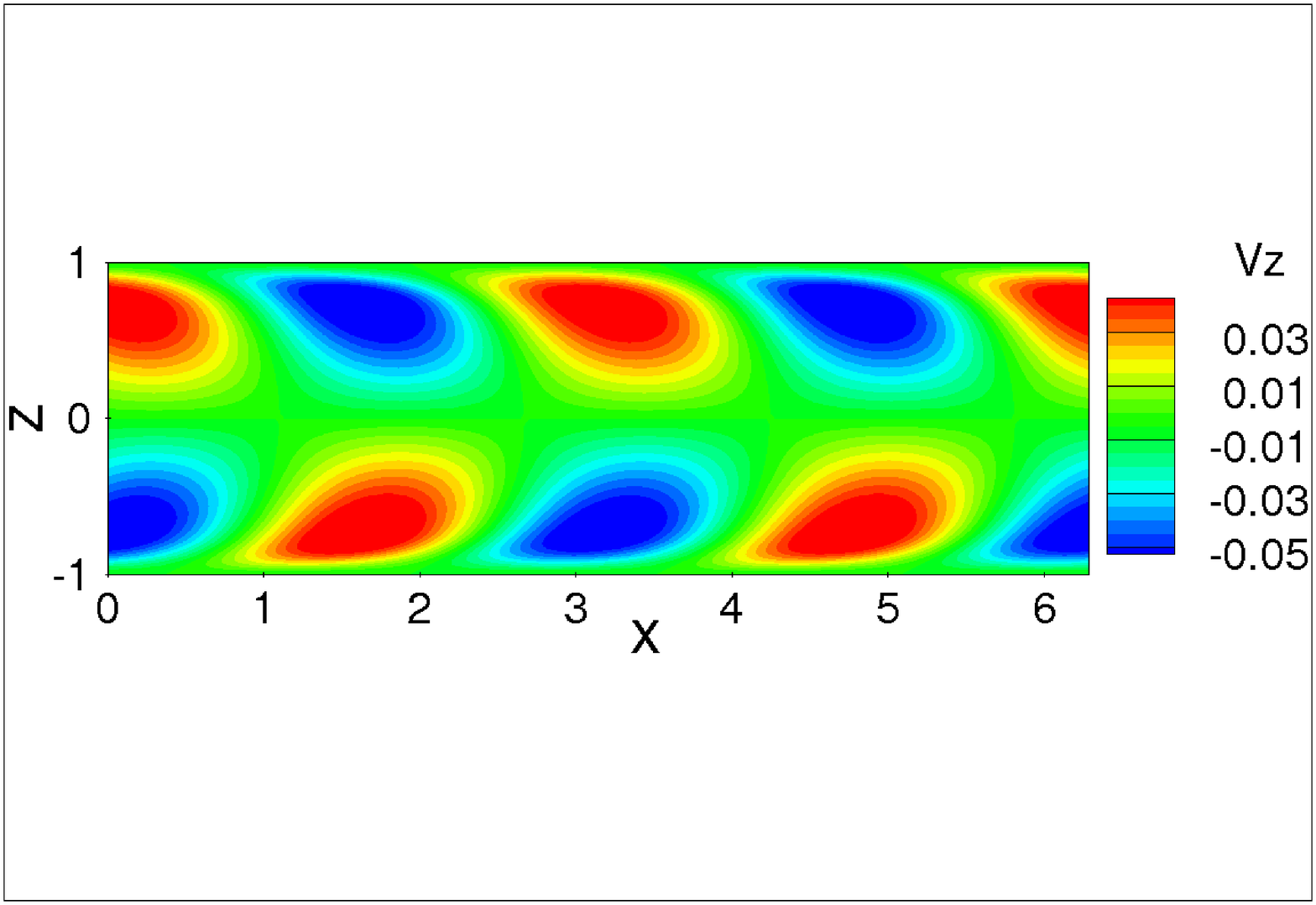}
                \label{fig:t161vxy}
        }
        \subfigure[]{
                \includegraphics[width=0.48\textwidth,trim=2mm 50mm 2mm 40mm,clip]{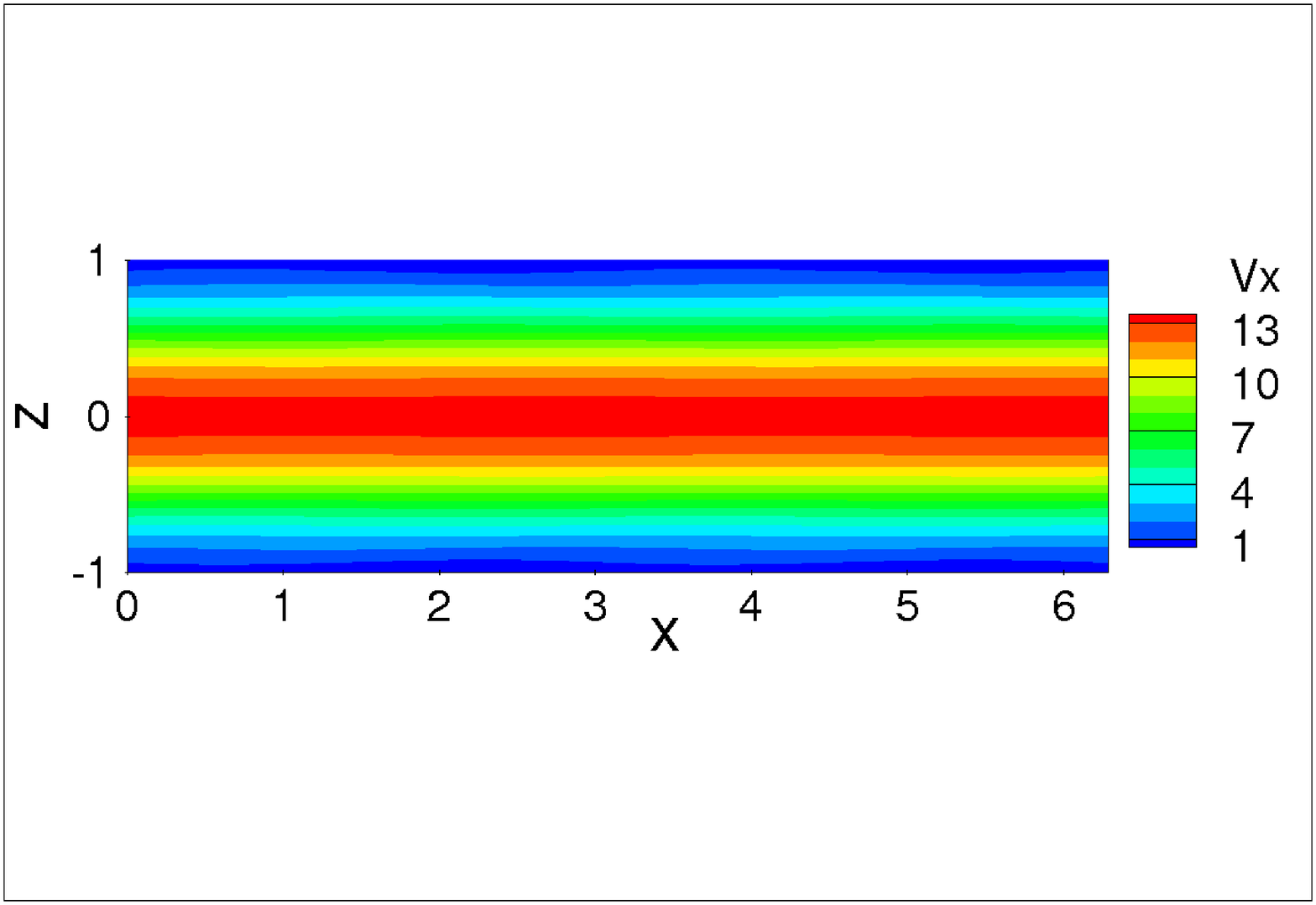}
                \includegraphics[width=0.48\textwidth,trim=2mm 50mm 2mm 40mm,clip]{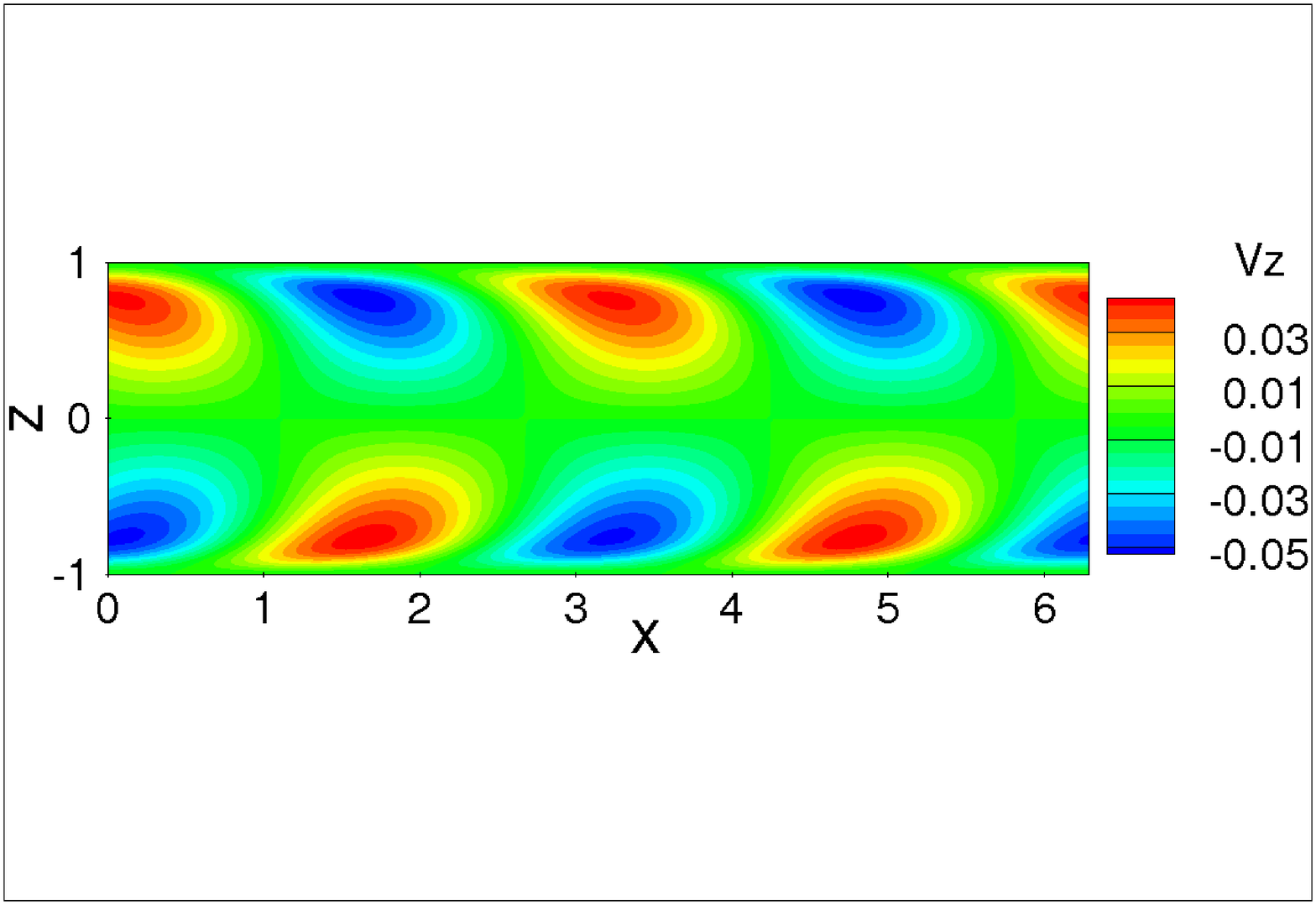}
                \label{fig:t601vxy}
        }
\caption{ Contours for the evolution of velocity field to the Poiseuille-like state as the flow accelerates from rest. (a) $t=0$, (b) $t=16.1$, (c) $t=60.1$. 
Parameters are $Q=0.5$, $\beta=1$, $\kappa=1$ and $\epsilon=5\times10^{-3}$. Left column: streamwise velocity $v_{x}$. Right column: 
wall-normal velocity $v_{z}$.}
\label{fig:vxvy}
\end{figure}
severing or reconnection of field lines are observed in this regime.
\begin{figure}[]
        \subfigure[]{
                \includegraphics[width=1.0\textwidth,trim=10mm 60mm 7mm 60mm,clip]{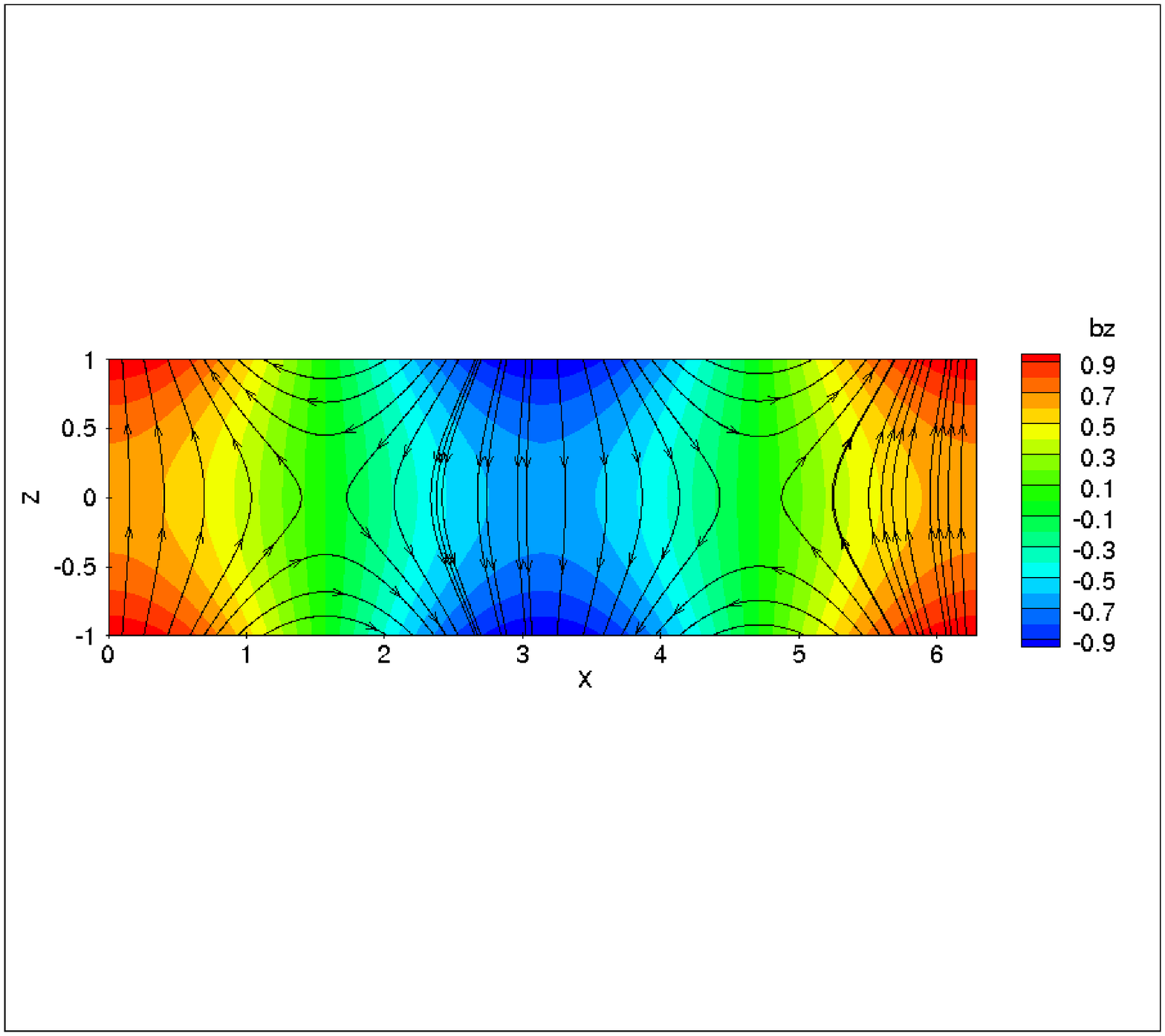}
                \label{fig:t0_flines}
        }
        \subfigure[]{
                \includegraphics[width=1.0\textwidth,trim=10mm 60mm 7mm 60mm,clip]{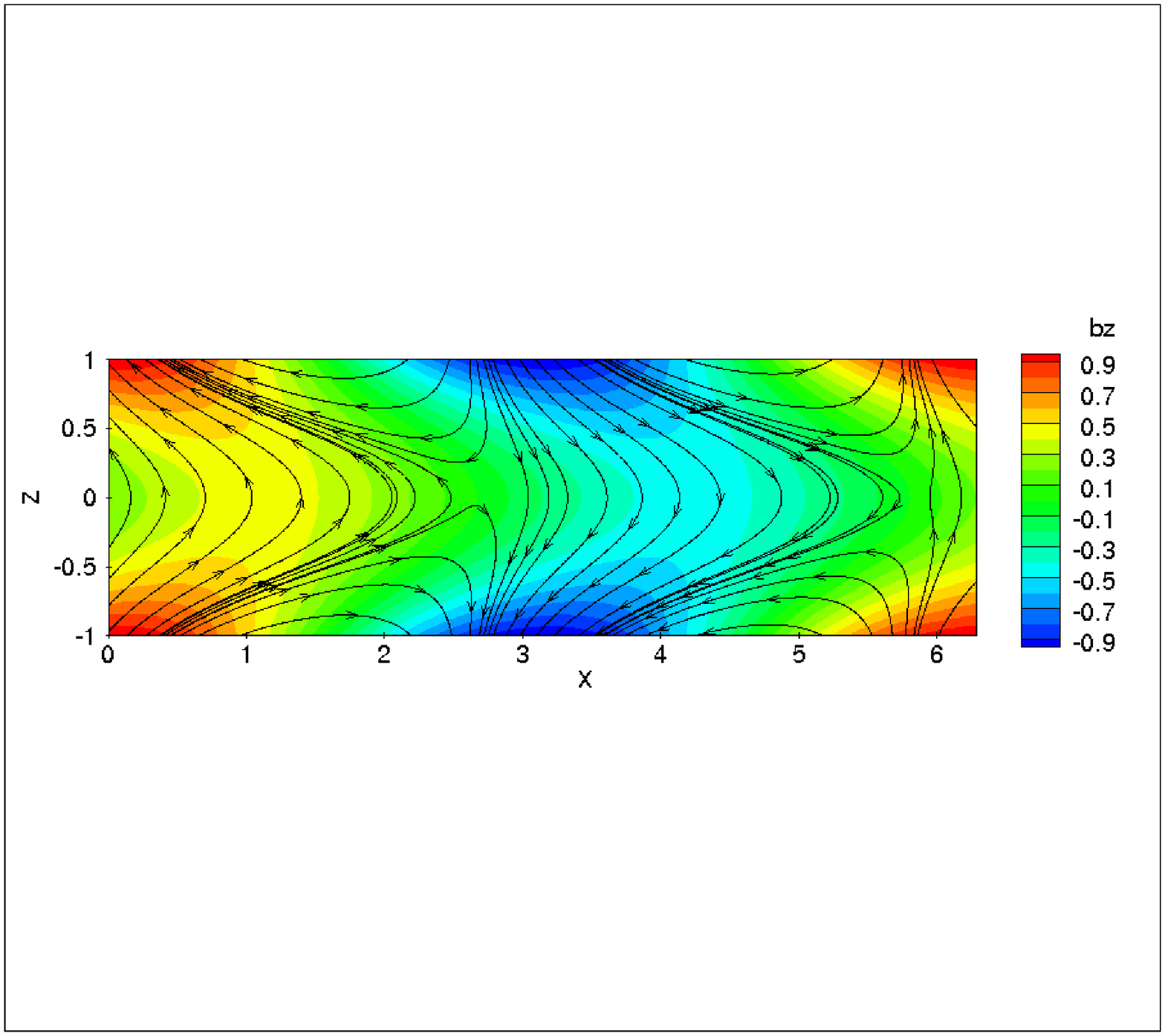}
                \label{fig:t161_flines}
        }
        \subfigure[]{
                \includegraphics[width=1.0\textwidth,trim=10mm 60mm 7mm 60mm,clip]{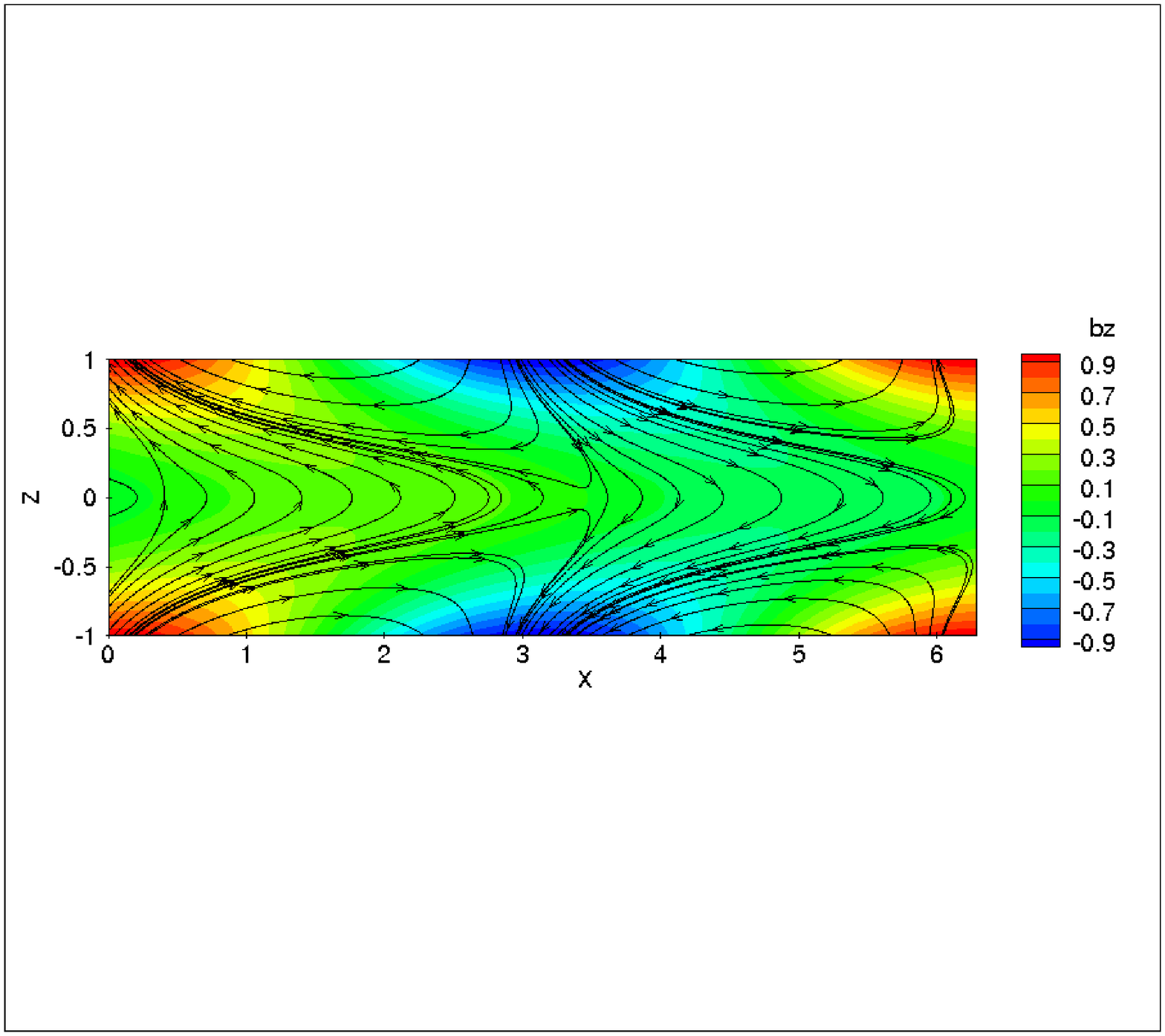}
                \label{fig:t601_flines}
        }
\caption{ Advection and expulsion of magnetic flux as the flow starts to accelerate from rest. (a) $t=0$, (b) $t=16.1$, (c) $t=60.1$. Parameters are $Q=0.5$, $\beta=1$, 
$\kappa=1$ and $\epsilon=5\times10^{-3}$. Contours coloured by $b_{z}$.}
\label{fig:bflines}
\end{figure}
\begin{figure}[]
        \subfigure[]{
                \includegraphics[width=0.35\textwidth,trim=55mm 60mm 70mm 60mm,clip]{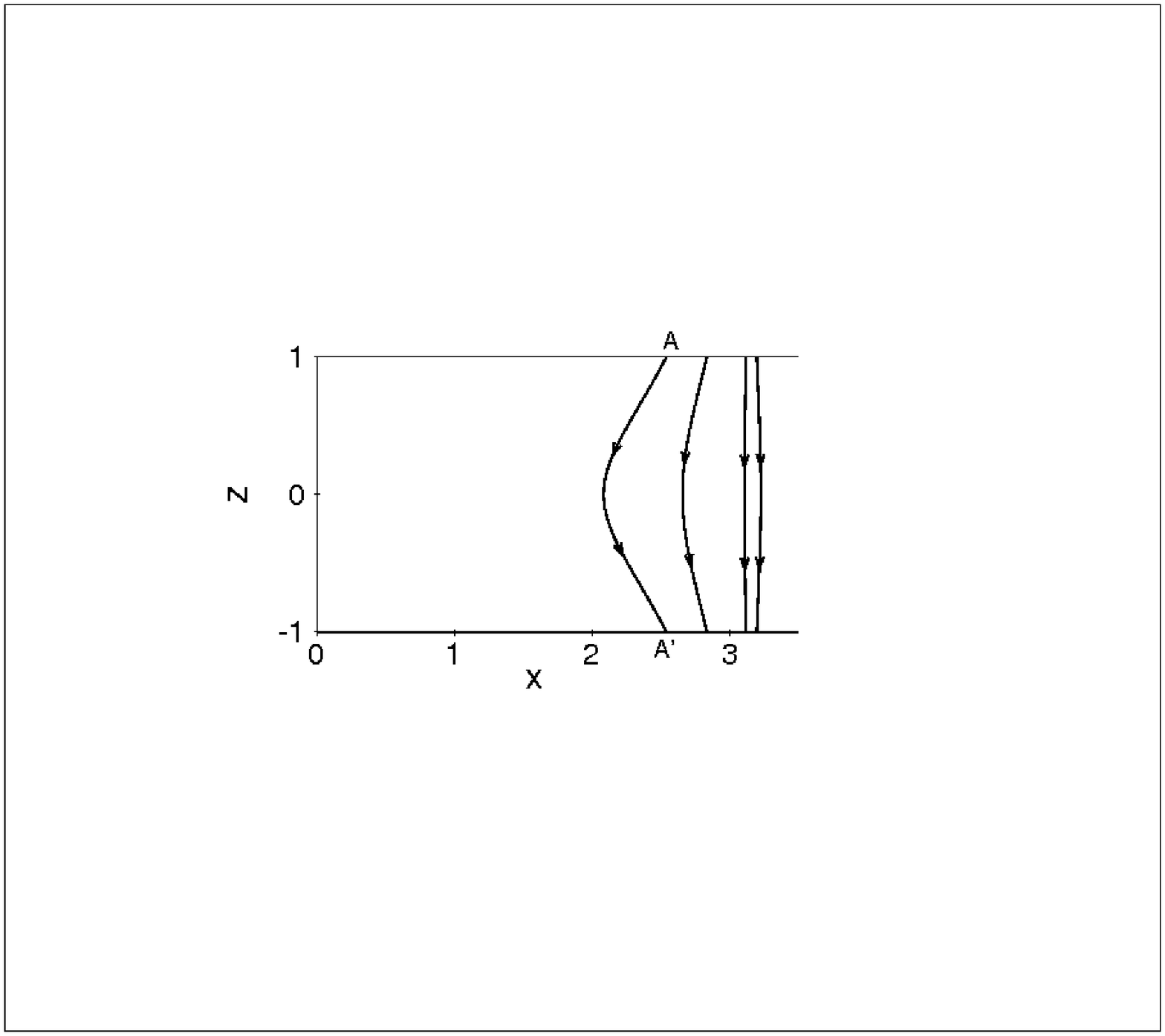}
                \label{fig:t0}
        }
        \subfigure[]{
                \includegraphics[width=0.35\textwidth,trim=55mm 60mm 70mm 60mm,clip]{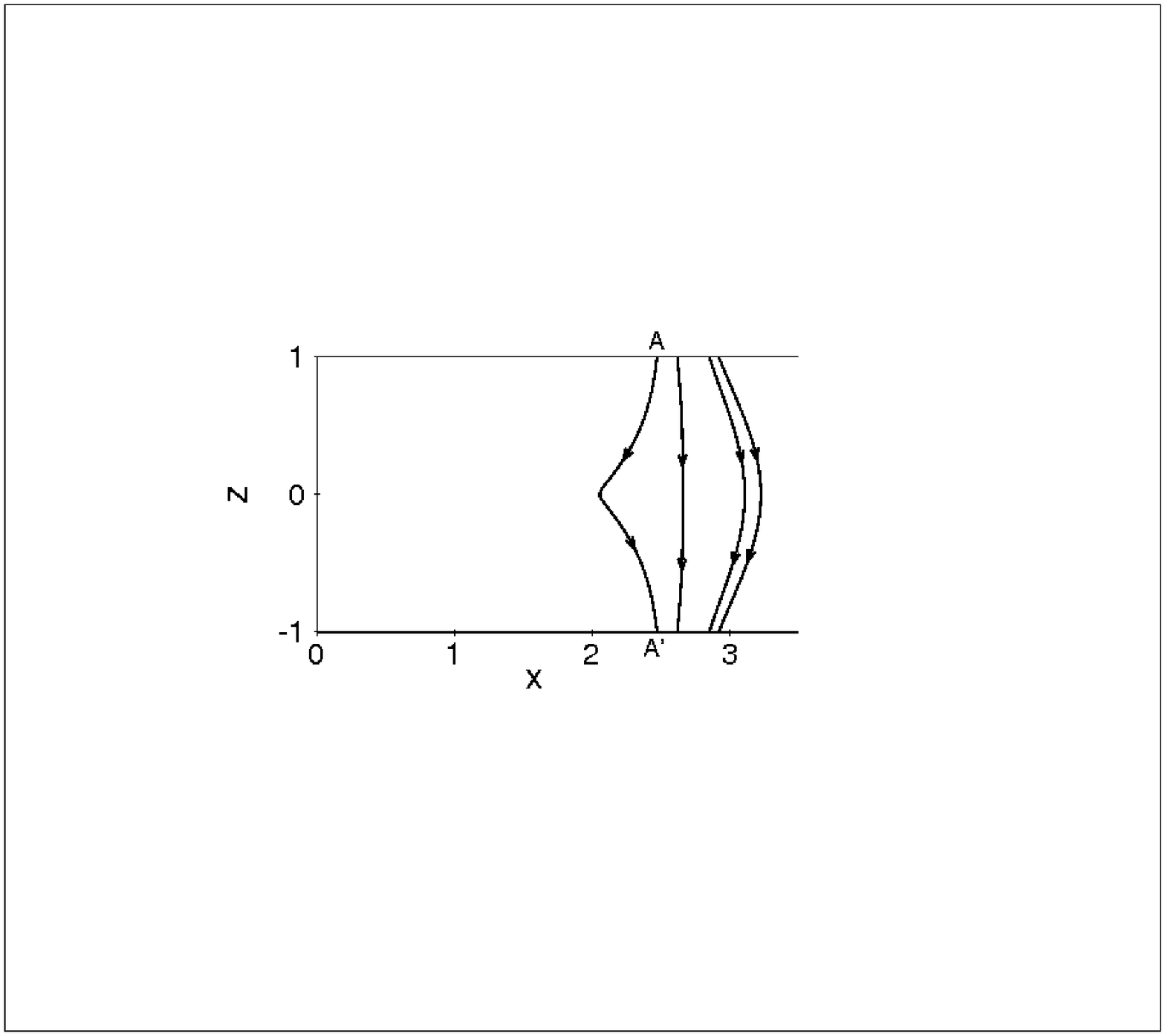}
                \label{fig:t20}
        }
        \subfigure[]{
                \includegraphics[width=0.35\textwidth,trim=55mm 60mm 70mm 60mm,clip]{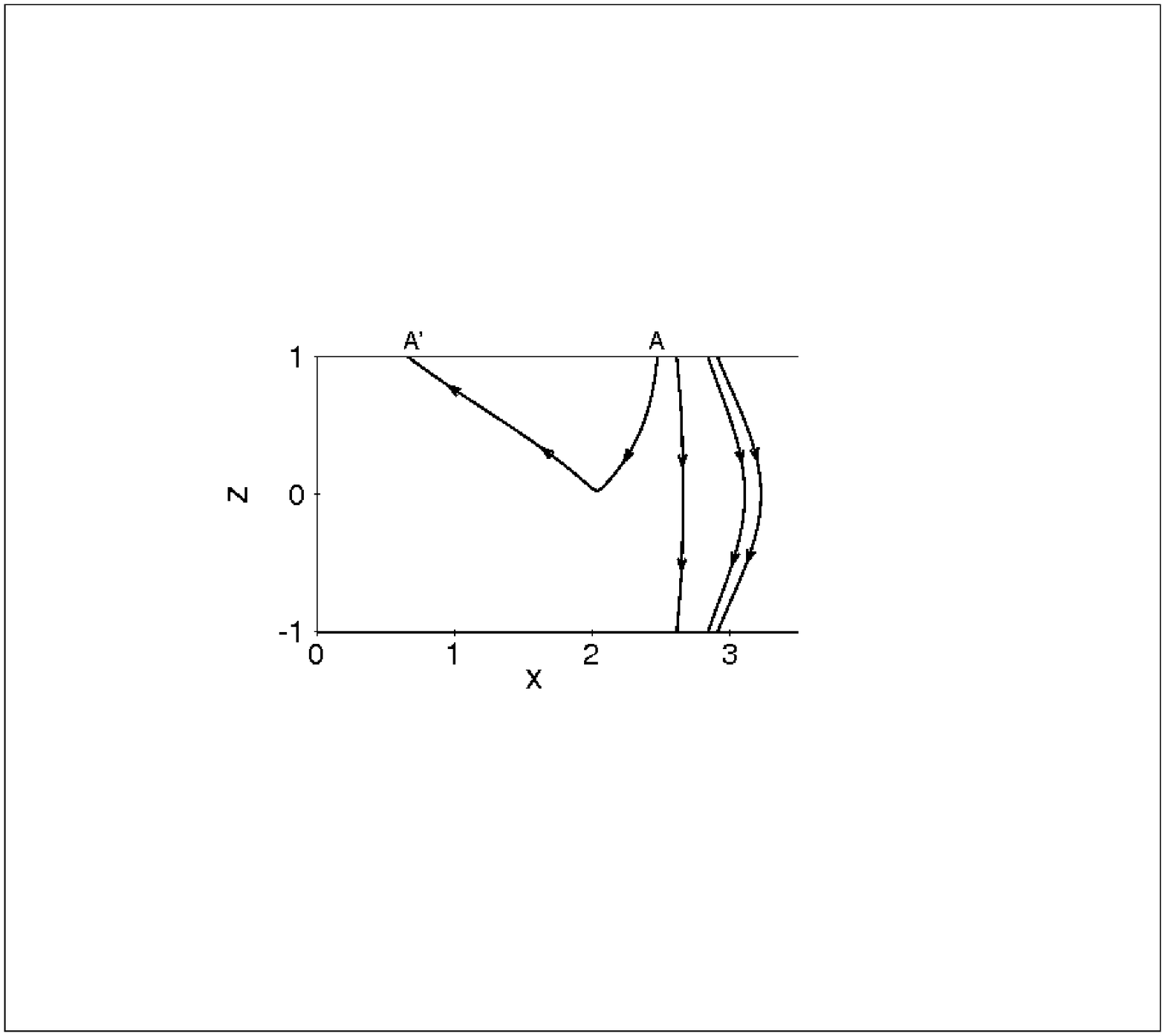}
                \label{fig:t21}
        }
        \subfigure[]{
                \includegraphics[width=0.35\textwidth,trim=55mm 60mm 70mm 60mm,clip]{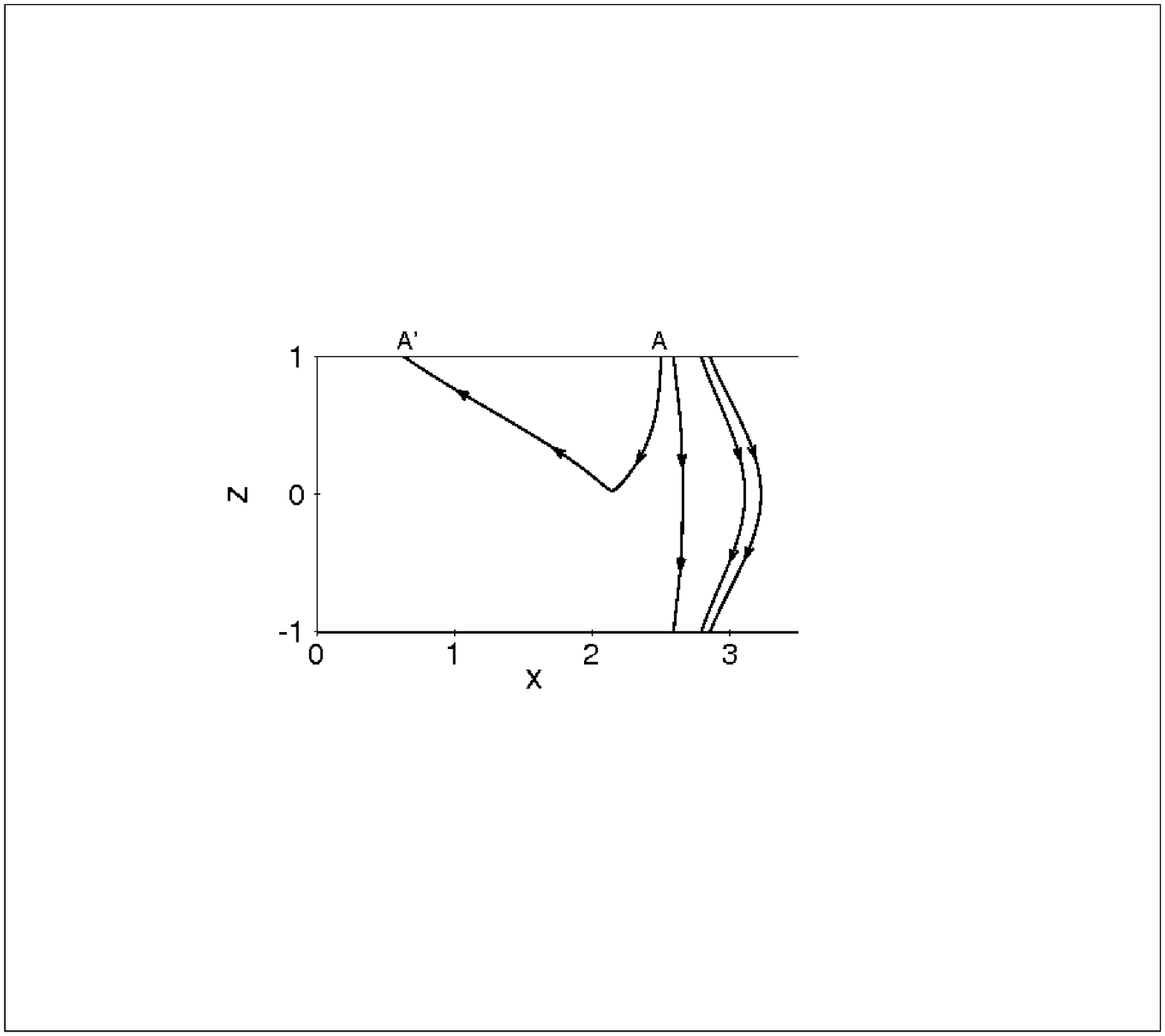}
                \label{fig:t29}
        }
        \subfigure[]{
                \includegraphics[width=0.35\textwidth,trim=55mm 60mm 70mm 60mmm,clip]{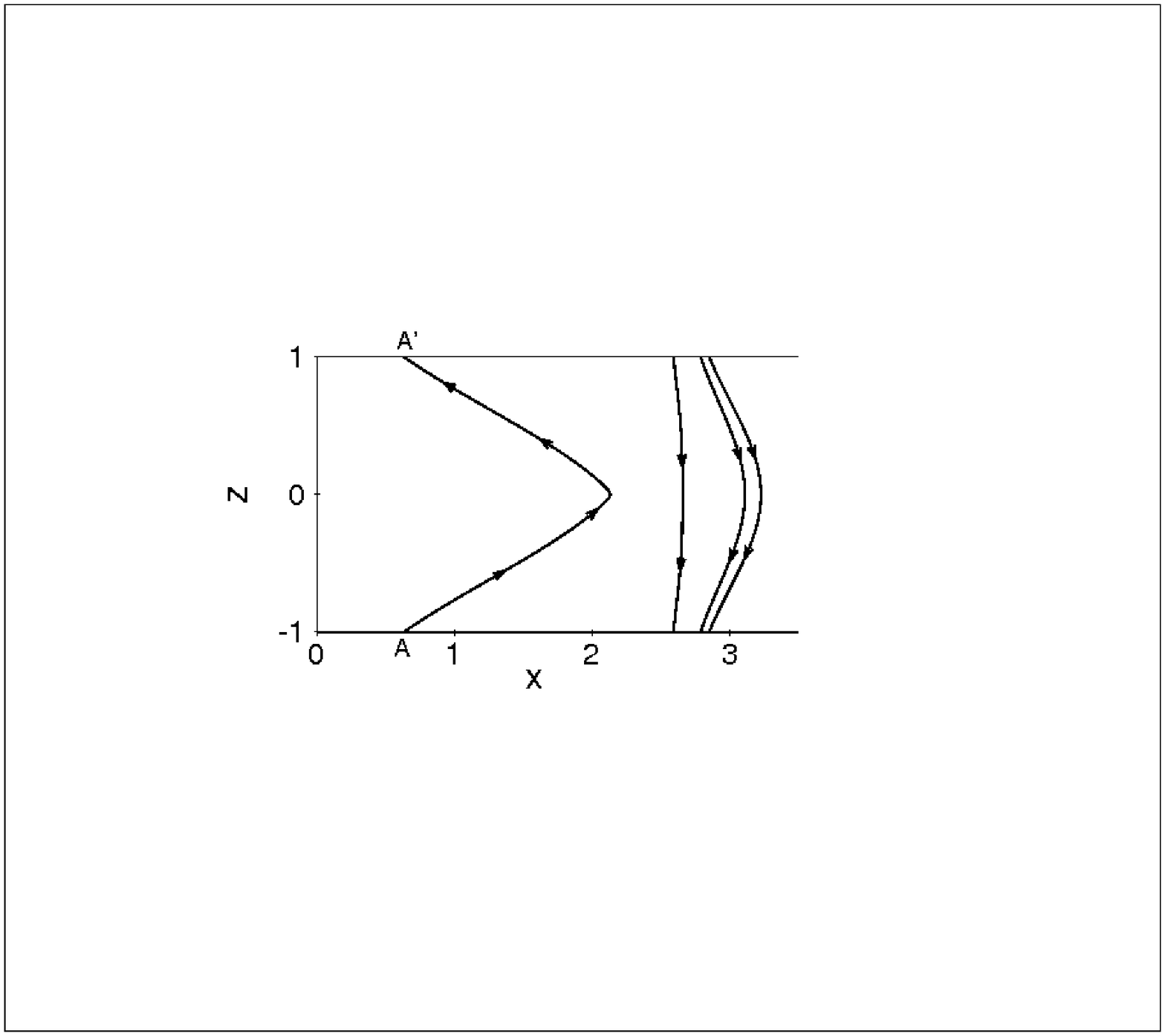}
                \label{fig:t30}
        }
        \subfigure[]{
                \includegraphics[width=0.35\textwidth,trim=55mm 60mm 70mm 60mm,clip]{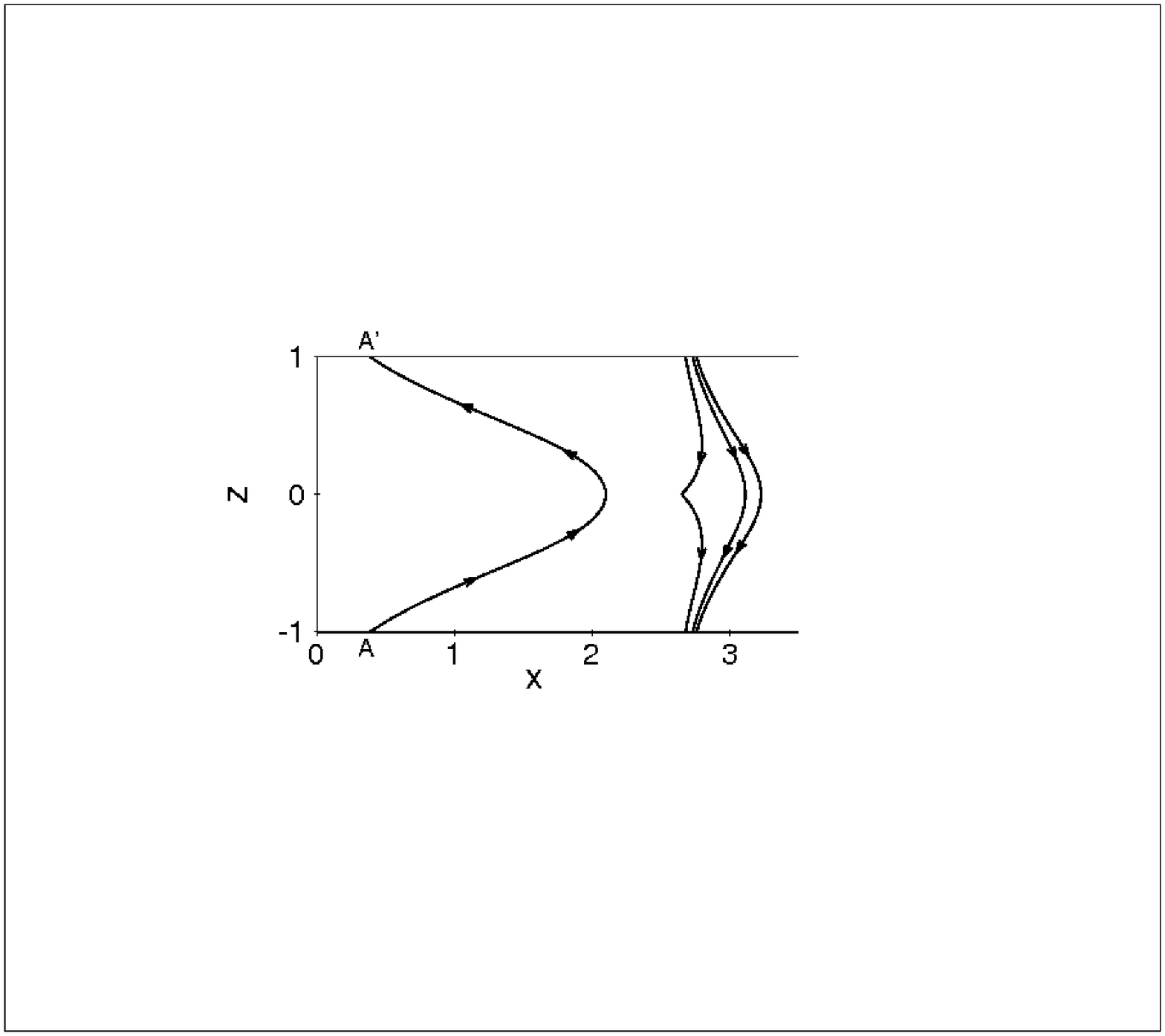}
                \label{fig:t154}
        }
        \subfigure[]{
                \includegraphics[width=0.35\textwidth,trim=55mm 60mm 70mm 60mm,clip]{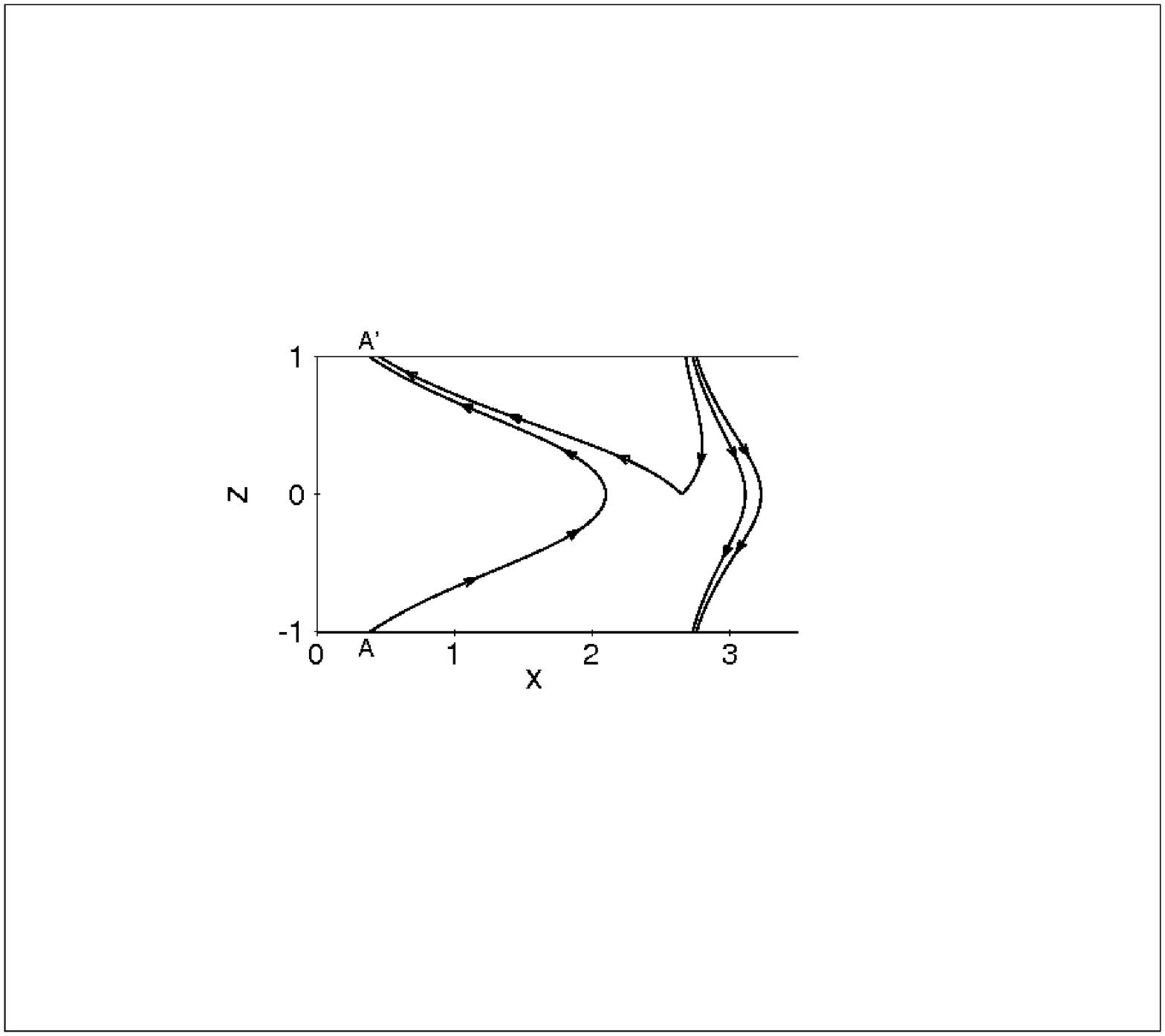}
                \label{fig:t155}
        }
        \quad \quad \quad \quad \quad \quad \quad \quad \quad \quad \quad \quad
        \subfigure[]{
                \includegraphics[width=0.35\textwidth,trim=55mm 60mm 70mm 60mm,clip]{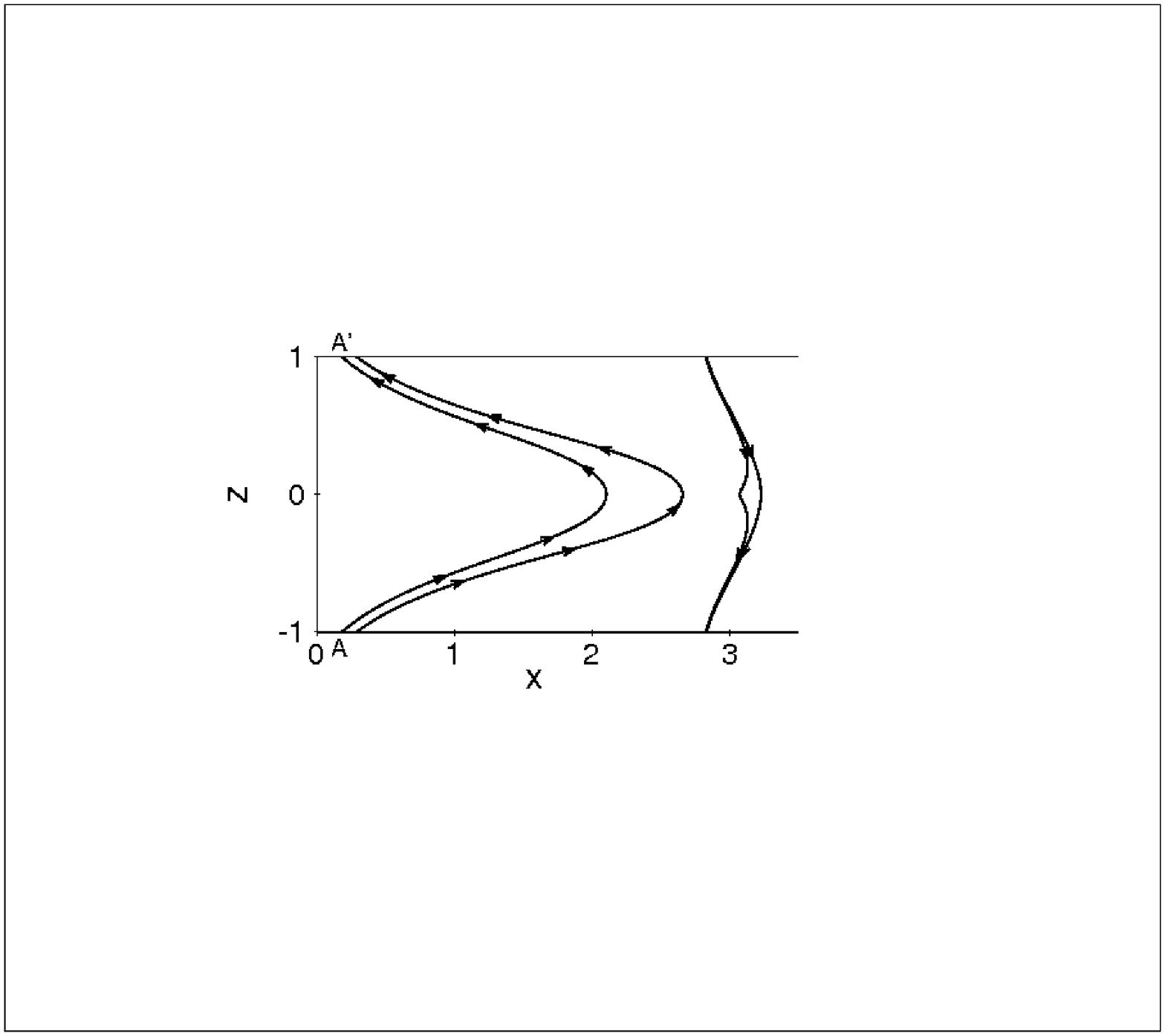}
                \label{fig:t370}
        }
\caption{ Severing and reconnection of magnetic field lines shown in a small section of the channel when the flow accelerates from rest.
(a) $t=0$, (b) $t=2.0$, (c) $t=2.1$, (d) $t=2.9$, (e) $t=3.0$, (f) $t=15.4$, (g) $t=15.5$, (h) $t=37.0$. Parameters are $Q=0.5$, $\beta=1$, $\kappa=1$ and $\epsilon=5\times10^{-3}$.
Following the line marked AA' shows a characteristic reconnection and stretching pattern that it undergoes leading to the expulsion of magnetic flux in the core.}
\label{fig:snapping}
\end{figure}
\subsection{Poiseuille regime}
The Poiseuille regime is a result of the runaway effect, where there is a significant acceleration of the flow due to considerable bending and severing of field lines and subsequent 
expulsion of magnetic flux. The streamwise velocity profiles show significant gradients in the wall normal direction and hence look `Poiseuille-like'.
 It is observed that the flow
exhibits strongly unsteady behaviour even in the final (steady on average) state. Of particular interest is the initial phase of the transient flow that ensues when the flow starts to
accelerate from rest, due to the applied mean pressure gradient.
The evolution of the velocity field leading to almost complete expulsion of magnetic flux in the core ($z=0$) in such
a case is shown in Fig.~\ref{fig:vxvy} through snapshots of velocity component contours. It can be seen that the gradual acceleration of the streamwise velocity is accompanied by
relatively small wall normal velocity component on both side sides of the core, in a staggered arrangement. At the same time, the normal component of the magnetic field (which leads to
streamwise Lorentz force) in the core is gradually destroyed, due to the bending of the vertical field lines (due to advection) as well as the reconnection of the field lines, depending on the streamwise location in the channel.
Reconnection here refers to the rearrangement of magnetic field line topology. 
This can be observed in Fig.~\ref{fig:bflines}, where snapshots of the configuration of the 
magnetic field lines show the eventual decay of magnetic flux in the core and significant shifting (advection) of the X-points, due to
reconnection events (the mean flow is from left to right).
\begin{figure}[!h]
\includegraphics[width=0.95\textwidth]{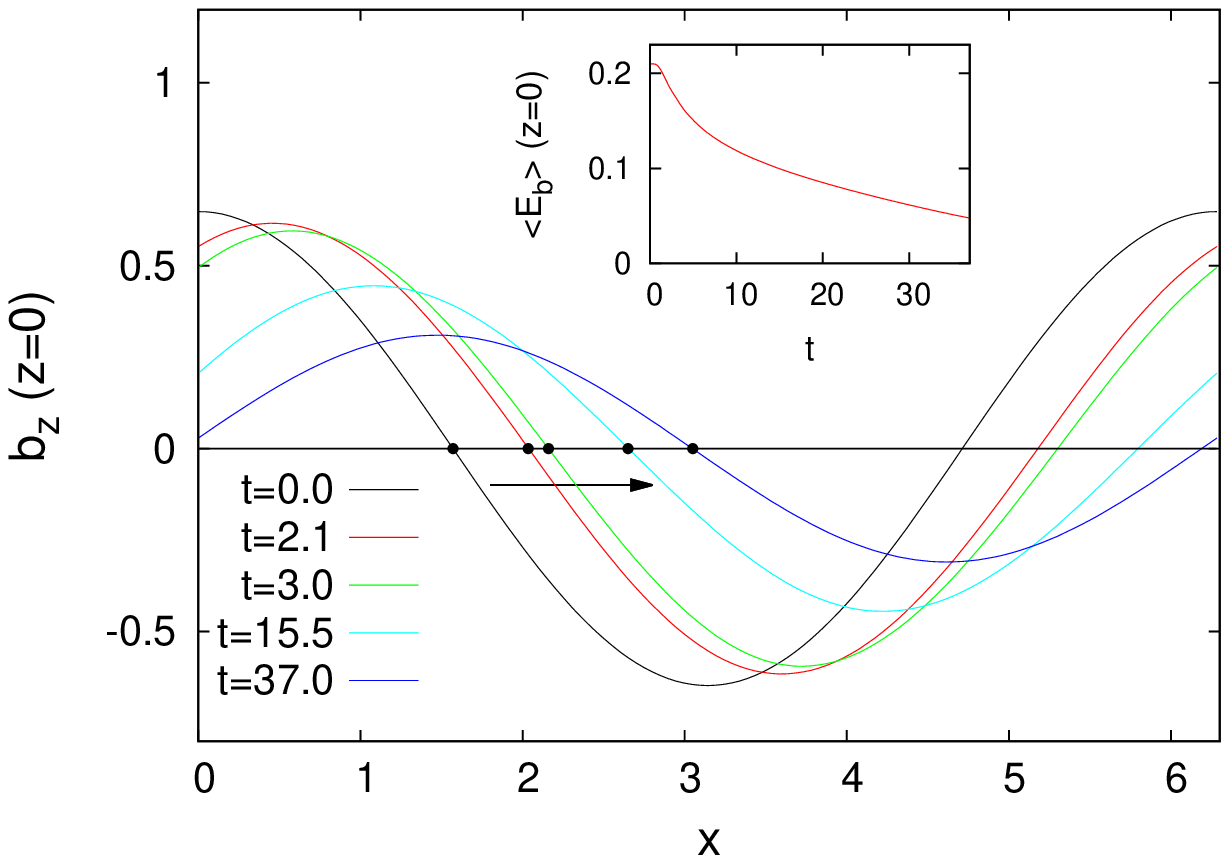}
\caption{Evolution of the wall normal component of the magnetic field ($b_{z}$) along the channel centerline $z=0$ when the flow accelerates from rest (obtained from DNS). 
Parameters are $\beta=1$, $Q=0.5$ and $\epsilon=5\times10^{-3}$. Inset shows corresponding time decay of the mean magnetic energy on the centerline $<E_{b}>$ at $z=0$. We mark the
magnetic X-points and their steady shift further downstream by reconnection events.}
\label{fig:bzofx}     
\end{figure}
A specific common pattern in the reconnetion of the magnetic field lines was observed during the process of dynamic runaway.
A field line in the region with a strong negative wall normal component $b_{z}$ (those that are approximately located at $2.5<x<\pi$ in Fig.~\ref{fig:t0_flines}), undergo a two-fold 
reconnection process and transform into a field line with positive $b_{z}$ (except at the core, where it is zero). A typical example of this is shown in more detail in Fig.~\ref{fig:snapping}, where
the field line (marked at the ends by A and A') corresponding to the imposed magnetic field initially develops a sharp `pinch' at $z=0$ (Fig.~\ref{fig:t20}) before a reconnection event leading to both ends 
of the line attached to the top wall (Fig.~\ref{fig:t21}).
\begin{figure}[]
        \subfigure[]{
                \includegraphics[width=1.0\textwidth,trim=2mm 60mm 2mm 30mm,clip]{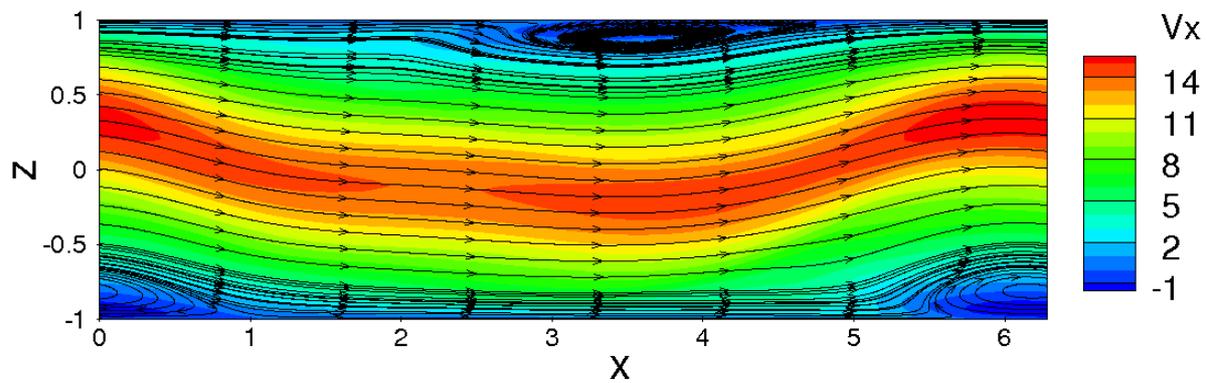}
                \label{fig:t03}
        }
        \subfigure[]{
                \includegraphics[width=1.0\textwidth,trim=2mm 60mm 2mm 30mm,clip]{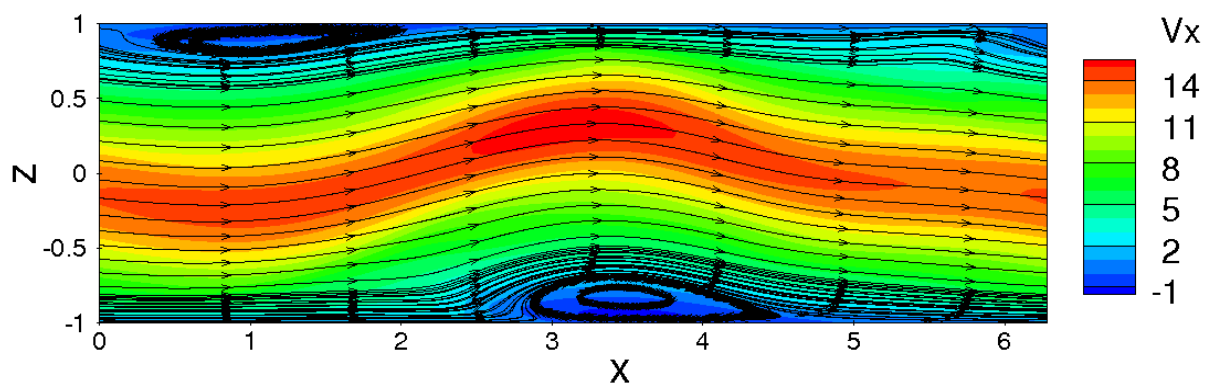}
                \label{fig:t11}
        }
        \subfigure[]{
                \includegraphics[width=1.0\textwidth,trim=2mm 60mm 2mm 30mm,clip]{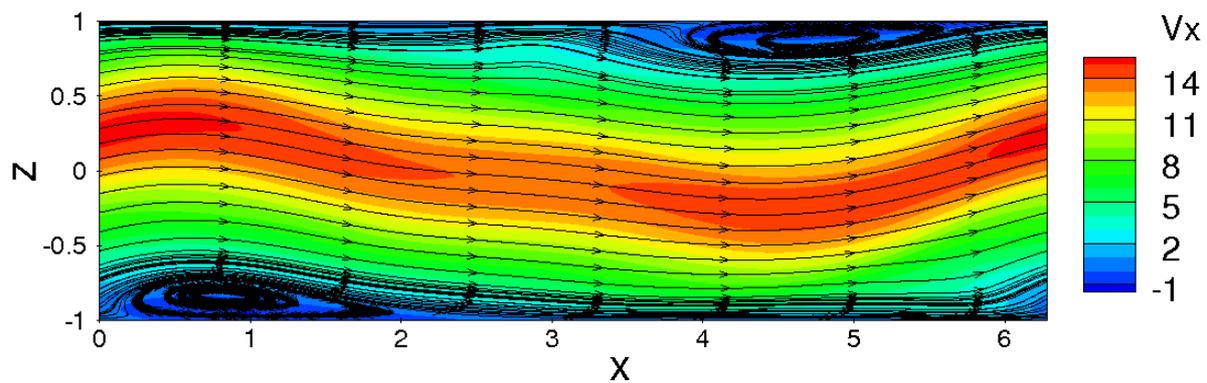}
                \label{fig:t19}
        }
\caption{ Snapshots of streamlines of the velocity field in the final (steady on average) state showing the transport of vortices near the wall. Coloured by contours of $v_{x}$.
(a) $t=0.3$, (b) $t=1.1$, (c) $t=1.9$. Parameters are $Q=0.5$, $\beta=1$, $\kappa=1$ and $\epsilon=5\times10^{-3}$.}
\label{fig:vlines_poiseuille}
\end{figure}
After some further stretching, another reconnection event occurs as seen from Fig.~\ref{fig:t29} to Fig.~\ref{fig:t30} leading to a reversal of the direction of the
magnetic field as compared to the initial state. Subsequently the field line is stretched significantly in the flow direction as shown from Fig.~\ref{fig:t29} 
through Fig.~\ref{fig:t370} leading 
to the final topology of the line that also results in $b_{z}=0$ at $z=0$. Interestingly, these series of events is seen to occur to every flux line (in the region considered, $2.5<x<\pi$)
 in a sequential manner from left to right. This is clearly seen from the pinching and reconnections occuring to the field line next to AA' seen from Fig.~\ref{fig:t154} 
through Fig.~\ref{fig:t370}.

The decay of the magnetic flux in the core during this period (from $t=0$ to $t=370$) is shown in Fig.~\ref{fig:bzofx}, where the evolution of $b_{z}(x)$ along the
channel centerline is plotted. Advection of X-points in the streamwise direction and the decay of the amplitude of $b_{z}$ can be clearly observed. This is accompanied by
the temporal decay of the mean magnetic energy at the channel centerline which defined as 
\begin{equation}
\label{eq:magenergy} <E_{b}>={l_{x}}^{-1}\int \limits_{0}^{l_{x}}\left(b_{x}^2+b_{z}^2\right)dx.
\end{equation}
This decay contributes to the growth of kinetic energy and hence the runaway process.

As mentioned previously, the final state occuring in the Poiseuille regime shows a strongly unsteady behaviour, with secondary flow structures which become more frequent at lower
viscosities. A typical example is shown in Fig.~\ref{fig:vlines_poiseuille}, where two large vortex structures are observed, one on either wall and seperated in the streamwise direction.
These vortices (or recirculation zones) are advected along the mean flow (Figs.~\ref{fig:t03} through \ref{fig:t19}), leading to an almost time-periodic behaviour of the flow. Absence of
chaotic states might be attributed to low Reynolds numbers and short domain length in the problem. 

\subsection{Comparison with the predictions of KM82}
\begin{figure}[]
        \subfigure[]{
                \includegraphics[width=0.5\textwidth]{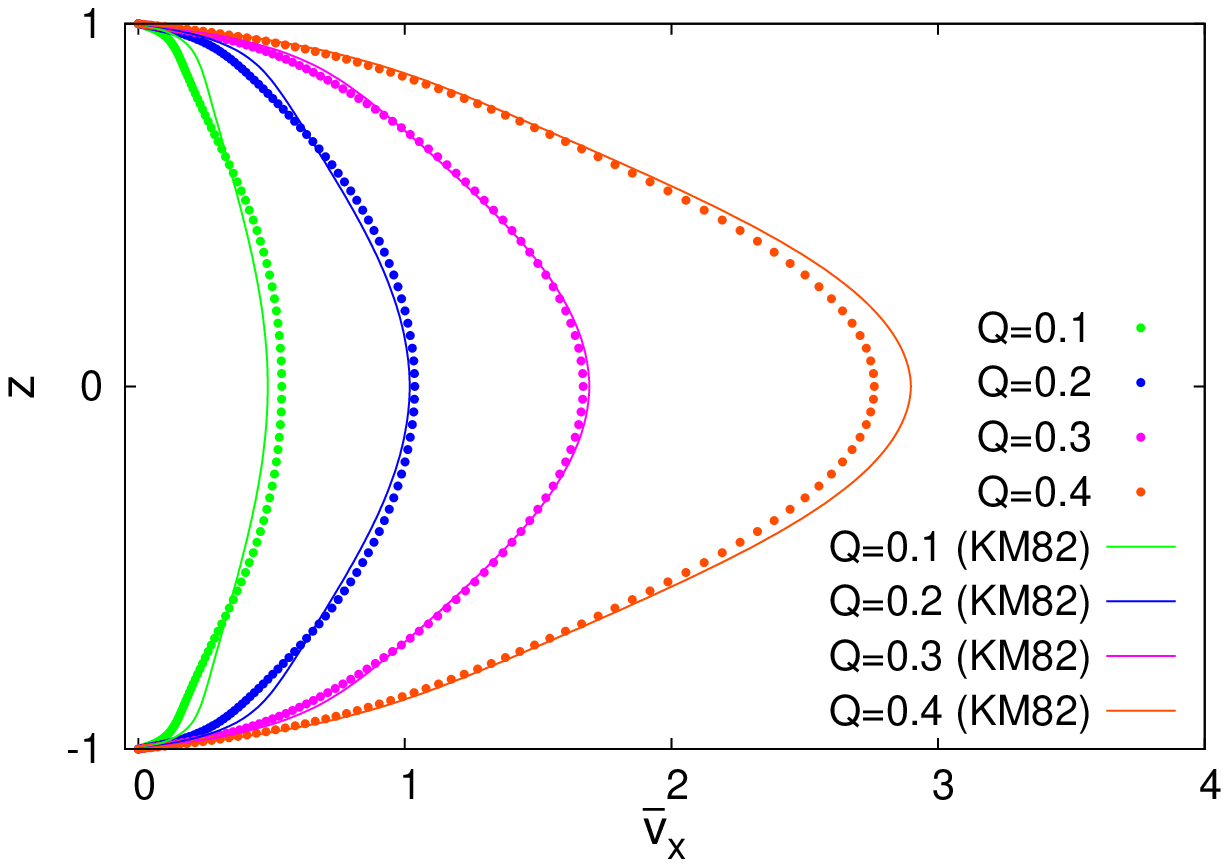}
                \label{fig:ucmeanvsq}
        }
        \subfigure[]{
                \includegraphics[width=0.5\textwidth]{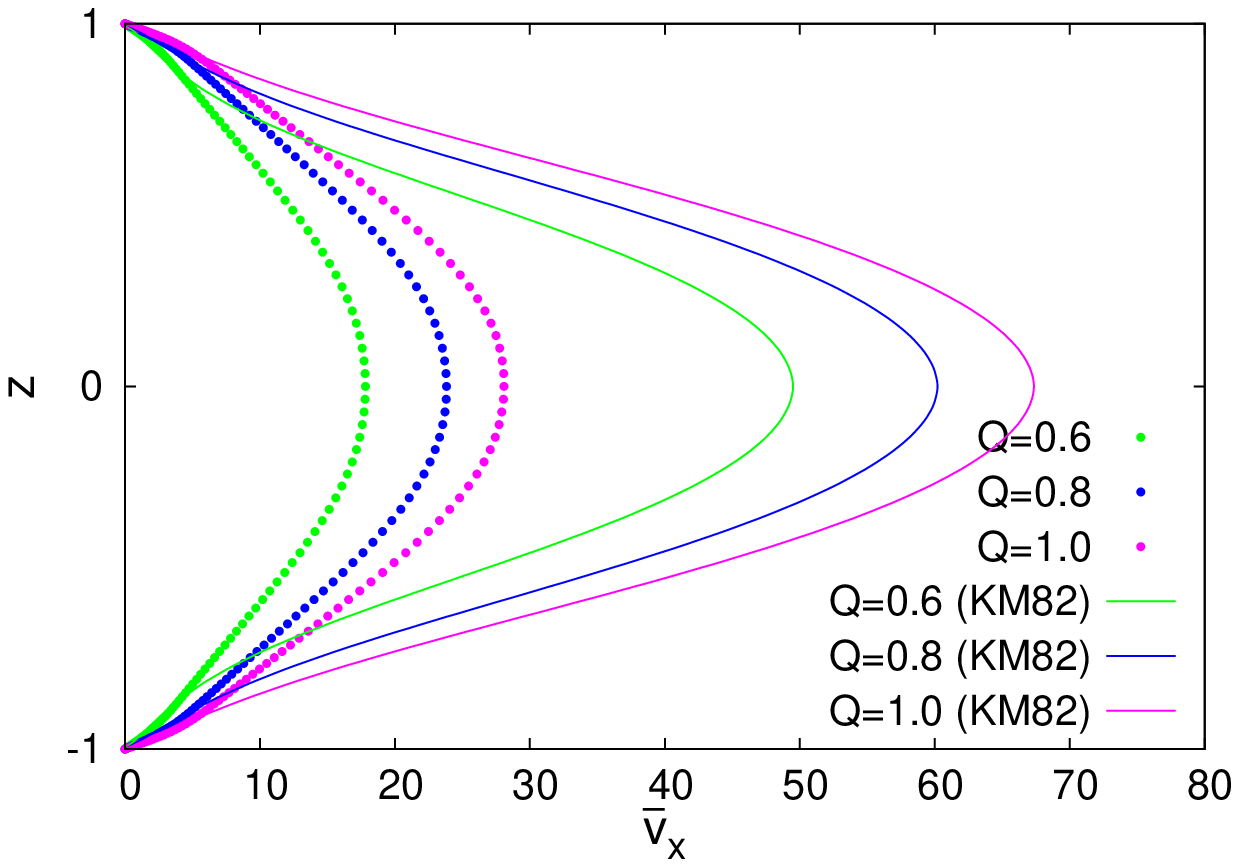}
                \label{fig:ucmeantvsq}
        }
\caption{(a) Steady state mean streamwise velocity profiles in the Hartmann regime, (b) Mean streamwise velocity profiles in the Poiseuille regime (streamwise and time averaged). Parameters are $\beta=1$, $\kappa=1$ and $\epsilon=5\times10^{-3}$.}
\end{figure}

We now turn to the comparison of DNS results with the model results of KM82 to investigate the validity of the model in predicting the steady states in both the regimes and also the 
location of bifurcation that leads to the transition between the two regimes. All the comparisons shown in this subsection correspond to $\epsilon=5\times10^{-3}$, $\beta=1$ and 
$\kappa=1$ with a channel length $L_{x}=2\pi$. Fig.~\ref{fig:ucmeanvsq} shows the mean streamwise velocity profiles in the Hartmann regime compared to the
prediction of KM82 at various values of the parameter $Q$. It can be observed that the model is accurate in this regime in terms of the magnitude of axial velocity although some 
differences can be seen in the shape of the profiles. 

However this is in contrast to the behaviour in the Poiseuille regime (see Fig.~\ref{fig:ucmeantvsq}), where 
the model strongly underpredicts the axial velocity $v_{x}$ and significant differences are observed in the shape of the velocity profiles. These differences can be attributed to the
effect of non-linearity which is more pronounced in the Poiseuille regime and the fact that the non-linear terms are neglected in the model equations of KM82. It must be pointed out
that in the case of Poiseuille regime, as the final state of the flow is strongly time-dependent, the axial velocity profiles from the DNS are obtained by time and streamwise averaging.

\begin{figure}[]
        \subfigure[]{
                \includegraphics[width=0.5\textwidth]{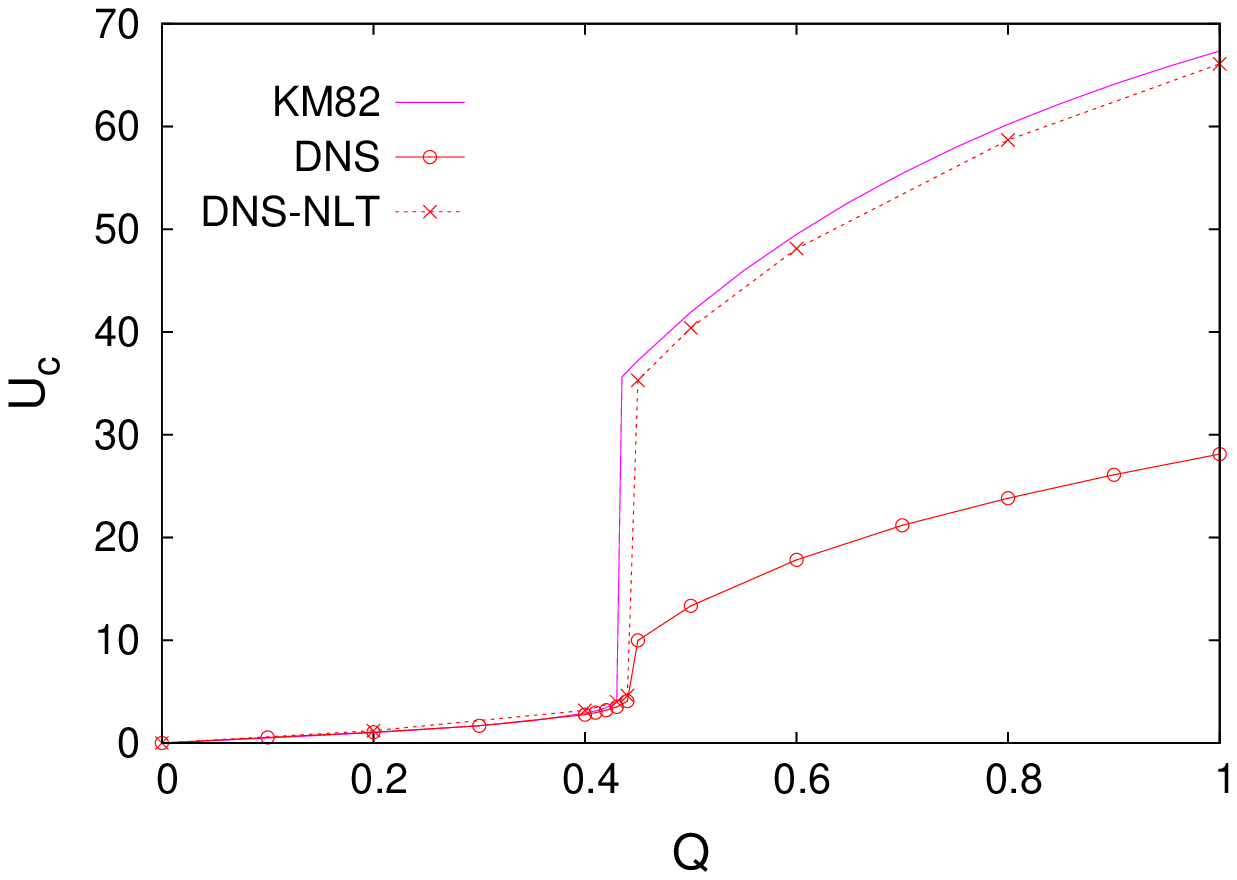}
                \label{fig:onlt}
        }
        \subfigure[]{
                \includegraphics[width=0.5\textwidth]{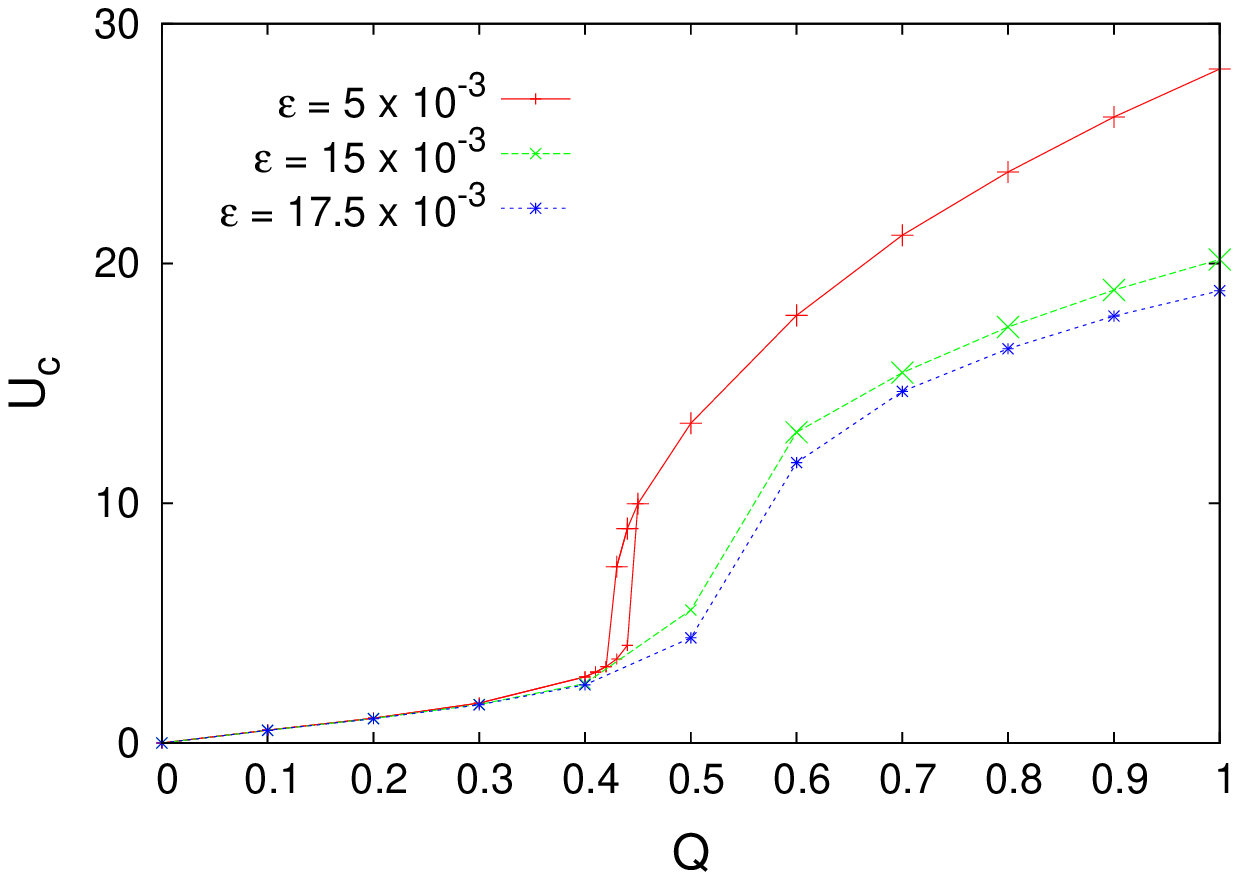}
                \label{fig:ucvsq}
        }
\caption{(a) Comparison of DNS results with the model predictions for $\epsilon=5\times10^{-3}$, $\beta=1$ and $\kappa=1$. The dotted line (denoted as DNS-NLT) indicates results obtained from
DNS by excluding the non-linear term., (b) Streamwise core velocities $U_{c}$ as a function of $Q$ obtained from DNS for various values of $\epsilon$. Dotted lines indicate that no 
hysteresis is observed. Parameters are $\beta=1$ and $\kappa=1$.}
\end{figure}

The dependance of the core axial velocity ($U_{c}=v_{x}$ at $z=0$) on $Q$ is shown in Fig.~\ref{fig:onlt}. The bifurcation from the Hartmann regime to the Poiseuille regime is observed
at $Q\sim0.43$, which is very close to that predicted by KM82 and the shape of the curve is in close match. The fact that non-linearity leads to the differences is confirmed through a
simulation that we performed dropping out the non-linear term ($\frac{1}{\beta\kappa}(\bm{v}\cdot \nabla)\bm{v}$) in \eqref{navierstokes}. Fig.~\ref{fig:onlt} shows that the curve 
obtained from DNS without the non-linear term tends very close to that of KM82.

\subsection{Effect of parameters $\epsilon$, $\beta$ and $\kappa$}
In this subsection, we present the effect of the parameters $\epsilon$, $\beta$ and $\kappa$ on the nature and location of the bifurcation along with the magnitude of core velocity
in the final state. Fig.~\ref{fig:ucvsq} shows $U_{c}$ versus $Q$ at different values of $\epsilon$. It is clear that at higher values of $\epsilon$ or lower Reynolds numbers, the transition
from the Hartmann to Poiseuille regimes does not show a distict jump but rather occurs in a continuous manner. In the parameter space with a clear bifurcation, the value of $Q$ at 
which the jump occurs is almost independent of the hydrodynamic Reynolds number (or $\epsilon$). In addition, when $\epsilon$ is low, a two-valued solution or hysteresis is 
observed near the bifurcation point (e.g. near $Q\sim0.43$ for $\epsilon=5\times10^{-3}$), depending on the direction in which the steady state is approached. Such a hysteresis effect
was also predicted by KM82.
 
The effect of $\epsilon$ on the core velocity ($U_{c}$) is
negligible in the Hartmann regime but is strong in the Poiseuille zone. All these observations are akin (qualitatively) to the predictions of KM82. 
\begin{figure}[]
        \subfigure[]{
                \includegraphics[width=0.5\textwidth]{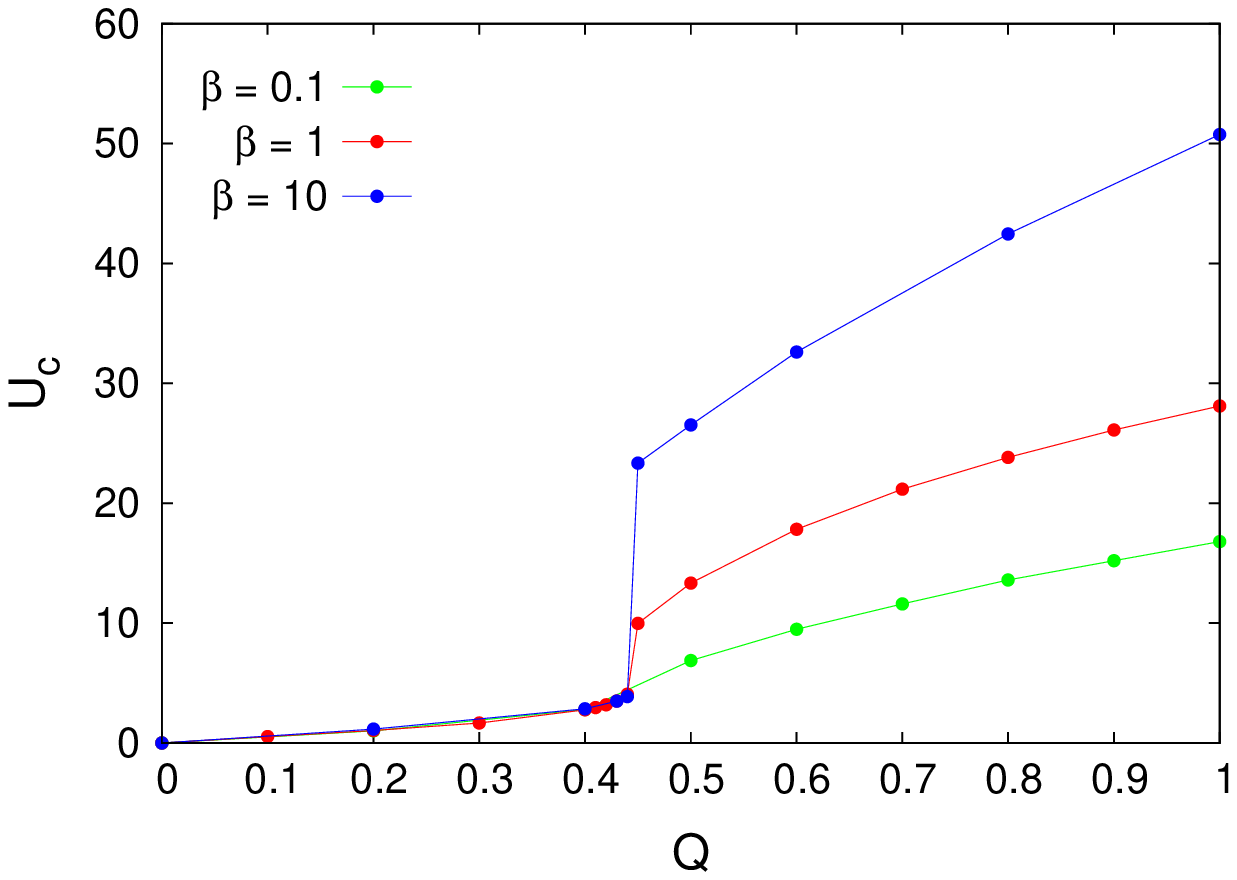}
                \label{fig:betavar}
        }
        \subfigure[]{
                \includegraphics[width=0.5\textwidth]{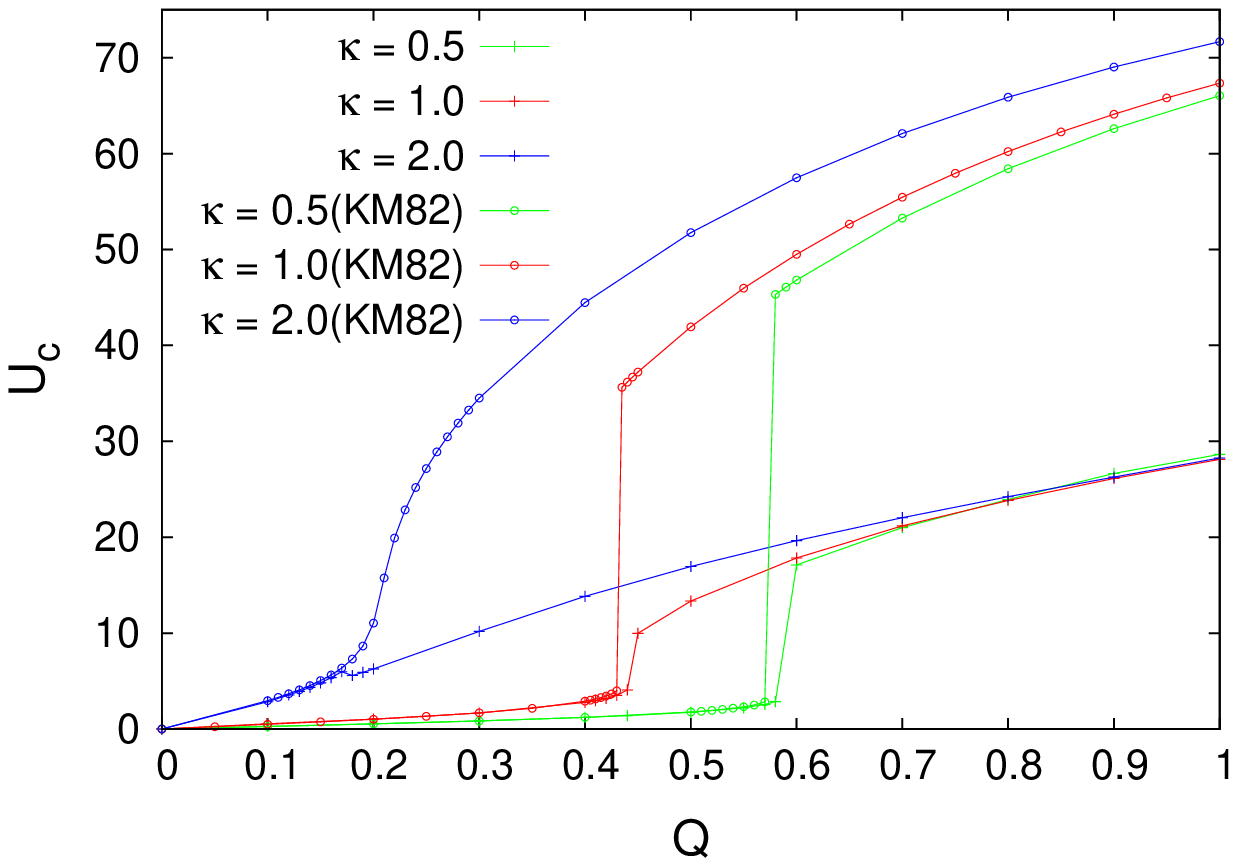}
                \label{fig:kappavar}
        }
\caption{(a) Effect of variation of magnetic Reynolds number $\beta$ in the $U_{c}$-$Q$ plane, obtained from DNS. Parameters are $\kappa=1$ and $\epsilon=5\times10^{-3}$. (b) Effect of
 variation of streamwise wavenumber $\kappa$ of the imposed magnetic field, obtained from DNS and KM82. Parameters are $\beta=1$ and $\epsilon=5\times10^{-3}$.}
\end{figure}
Interestingly, the effect of the magnetic Reynolds number $\beta$ on the $U_{c}$-$Q$ curve is very similar to that of the hydrodynamic Reynolds number, with higher levels of flux 
expulsion and core flow occuring at higher values of $\beta$, as can be seen in Fig.~\ref{fig:betavar}. It is interesting to note that steady state solutions of KM82 are independent of
the magnetic Reynolds number due to the association of $\beta$ with only the non-linear term (which is neglected in the model) in the momentum balance, as can be seen from 
equations \eqref{navierstokes} and \eqref{navierstokesk82}. Furthermore, the location of the bifurcation to the Poiseuille regime is almost
unaffected by variations in $\epsilon$ or $\beta$.
In contrast, the jump is observed to be very sensitive to the streamwise wavenumber ($\kappa$) of the imposed magnetic field $\bm{b}_{0}$. This is shown in 
Fig.~\ref{fig:kappavar}, indicating 
a clear increase in the value of $Q$ at which the bifurcation occurs and also the magnitude of the jump when $\kappa$ is decreased. This is very similar to the dependance on $\kappa$
predicted by the inviscid version of KM82, i.e. by using $\epsilon=0$ in equation \eqref{navierstokesk82}. In specific, for $\kappa=0.5$, inviscid KM82 predicts 
$Q_{c}\approx0.55$ as compared to $Q_{c}\approx0.59$ from viscous KM82 and $Q_{c}\approx0.6$ obtained from DNS with $\beta=1$ and $\epsilon=5\times10^{-3}$. 
At a higher wavenumber ($\kappa=2$), KM82 in the inviscid limit predicts a jump at a value of $Q{_c}=0.15$ whereas the viscous KM82 shows a continuous transition between the two
regimes. Interestingly in this case, DNS shows no clear demarcation between the Hartmann and the Poiseuille regimes, although a very small fall (rather than a jump) in $U_{c}$ 
is observed when $Q$ is increased from $0.17$ to $0.18$ and a corresponding onset of near wall recirculation zones at $Q=0.18$.

\section{Concluding remarks}
\label{sec:conclusions}
We presented results of direct numerical simulations of the dynamic runaway effect due to flux expulsion in a plane channel MHD flow. General features of the flow and magnetic
fields in the two regimes - the Hartmann and Poiseuille regimes - were studied. Our results show that the one-dimensional
model of Kamkar and Moffatt \cite{Kamkar:1982} is relatively accurate in the Hartmann regime and in the prediction of the location of the bifurcation to the Poiseuille regime (in the $U_{c}$-$Q$ plane).
However, significant differences in core velocity predictions are observed in the Poiseuille regime, attributed to the neglect of the non-linear terms in the Navier-Stokes equation.
The Poiseuille regime is seen to be strongly unsteady similar to that of travelling waves, but does not show spatial irregularity for the parameters that we discussed here. The location of the 
bifurcation is found to be
independent of the hydrodynamic and magnetic Reynolds numbers, but however is strongly affected by the wavenumber of the imposed magnetic field as lower
streamwise wavenumber ($\kappa$) leads to bifurcation to the Poiseuille regime at much higher values of $Q$ and vice versa. Finally, in contrast to the
model, the magnetic Reynolds number ($\beta$) has a substantial effect on the $U_{c}$-$Q$ curve, that is similar to the effect of the hydrodynamic Reynolds number ($\epsilon^{-1}$).

A full three dimensional DNS of this configuration to study well developed turbulence that might ensue in the Poiseuille regime can be a scope for future research. Also of interest
would be to quantify the rate of reconnection of magnetic field lines leading to flux expulsion in the core. This would also become much more demanding in three dimensions since the
field line topologies are more complex. 
Furthermore, in the case of electrically insulating walls which are of practical interest, magnetic boundary conditions could be non-trivial, as the secondary magnetic field 
transcends the channel walls and would require matching the exterior and interior magnetic fields at the wall boundaries.



\section{Acknowledgements}
We acknowledge financial support from the Deutsche Forschungsgemeinschaft within the framework of the Research Training Group 1567.\\

\bibliographystyle{elsarticle-num}
\bibliography{refs}
\biboptions{sort&compress}

\end{document}